\documentclass[10pt,journal,compsoc]{IEEEtran}

\ifCLASSOPTIONcompsoc
  \usepackage{amsmath}
  \newtheorem{theorem}{Theorem}[section]
  \newtheorem{lemma}{Lemma}[section]
   \newtheorem{definition}{Definition}[section]
  \usepackage{graphicx}
  \usepackage{tabularx}
  \usepackage{booktabs}
  \usepackage{epstopdf}
  \usepackage[nocompress]{cite}
  \usepackage{url}
  \usepackage{epstopdf}
  \usepackage{pifont}
  \usepackage{subfigure}
  \usepackage{amssymb}
  
\else
  \usepackage{cite}
  
\fi

\ifCLASSINFOpdf
\else
\fi
\begin{document}
%
\title{Soft Compression for Lossless Image Coding}
%
%
%
%

\author{Gangtao Xin,
        and Pingyi Fan,~\IEEEmembership{Senior Member,~IEEE}
        
\IEEEcompsocitemizethanks{\IEEEcompsocthanksitem Gangtao Xin and Pingyi Fan are with the Beijing National Research Center for Information Science and Technology and the Department of Electronic Engineering, Tsinghua University, Bejing 10084, China.\protect\\
E-mail: fpy@tsinghua.edu.cn
}
}

\IEEEtitleabstractindextext{%
\begin{abstract}
Soft compression is a lossless image compression method, which is committed to eliminating coding redundancy and spatial redundancy at the same time by adopting locations and shapes of codebook to encode an image from the perspective of information theory and statistical distribution. In this paper, we propose a new concept, compressible indicator function with regard to image, which gives a threshold about the average number of bits required to represent a location and can be used for revealing the performance of soft compression. We investigate and analyze soft compression for binary image, gray image and multi-component image by using specific algorithms and compressible indicator value. It is expected that the bandwidth and storage space  needed when transmitting and storing the same kind of images can be greatly reduced by applying soft compression.
\end{abstract}

}

\maketitle

\IEEEdisplaynontitleabstractindextext

%
\IEEEpeerreviewmaketitle

\IEEEraisesectionheading{\section{Introduction}\label{sec:introduction}}
\IEEEPARstart{T}{he} purpose of image compression is to reduce the number of bits required to represent an image as much as possible under the condition that the fidelity of the reconstructed image to the original image is higher than a required value. Image compression often includes two processes, encoding and decoding. Encoding is to convert the input image into binary code stream by a coding method, while the decoding is the reverse process of encoding, which aims to restore the original image from the binary code stream. 

There are two categories of image compression: lossy compression and lossless compression. Lossy compression allows the reconstructed image to be different from the original image, and it is supposed to be visually similar. Lossless compression requires the reconstructed image to be exactly the same as the original image, which leads to the compression ratio of lossless compression being much smaller than that of lossy compression. Although the compression ratio of lossy compression is higher, lossless compression still plays a gigantic role in a great deal of fields. Lossless compression should be employed in situations where errors cannot be tolerated or where there is significant value, such as medical images, precious cultural relics, deep space exploration, deep sea exploration and digital libraries. 

Supposing that we regard an image as a random process, the minimum expected codeword length per pixel $L_n^*$ will approach the entropy rate asymptotically, as shown in formula (\ref{Eq:Ln}). However, for an actual image, the upper and lower bound of formula (\ref{Eq:Ln}) cannot be calculated theoretically because one can't know the spatial correlation of pixels clearly. 
\begin{equation}
{H(X_1, X_2,..., X_n) \over n} \leq L_n^* < {H(X_1, X_2,..., X_n) \over n} + {1 \over n}
\label{Eq:Ln}
\end{equation}
where $H(X_1, X_2,..., X_n)$ is the joint entropy of the symbol series $\{X_i, i=1,2,...,n\}$.

It's impossible to reach the entropy rate and everything we can do is to make great efforts to get close to it. Soft compression \cite{9247990} was recently proposed, which uses locations and shapes to reflect the spatial correlation of an image as far as possible from the perspective of information theory and statistical distribution, so as to achieve a better compression effect.

In the literature, most of image compression methods mainly consider three aspects to reduce the required number of bits when representing an image: coding redundancy, spatial redundancy and irrelevant information. Coding redundancy refers to that the probability of each pixel value in an image is diverse, so the average length can be reduced from the perspective of coding. Spatial redundancy means that pixels are spatially related where a pixel is similar to or depends on adjacent pixels \cite{gonzalez2004digital}, so the repeated information can be omitted. Irrelevant information refers to that an image contains information irrelevant to the human visual system or purpose, which leads to the redundancy. Image compression techniques usually improve its compression effect from one or several aspects.

\subsection{Image Compression Method}
Huffman coding \cite{huffman1952a} is an extraordinary method to eliminate coding redundancy for a steam of data, and any other code for the same alphabet cannot have a lower expected length than the code constructed by Huffman coding \cite{cover1999elements}. Arithmetic coding \cite{5390830} and Golomb coding \cite{golomb1966run} are also approaches to eliminating coding redundancy, and they all require accurate probability models of input symbols. Run-Length coding\cite{1092283} represents runs of identical intensities by a new coding value and length, but it may result in data expansion when there are few runs of identical intensities \cite{gonzalez2004digital}. LZW coding \cite{welch1984technique} is a method to remove spatial redundancy, assigning fixed-length codewords to variable length sequences of source symbols, but it is easy to cause data explosion, especially when the input is with large size or irregular. Image predictive coding \cite{boon2000image} is a means of transforming spatial redundancy into coding redundancy through prediction error.

Transform coding \cite{4711713,8652455} maps an image from the spatial domain to transform domain, and then encodes the coefficients of transform domain to attain the compression effect. This method can reduce the irrelevant information in an image from the visual point of view. As a tool of multi-resolution analysis, wavelet coding \cite{136597,remya2012wavelet} has been widely concerned and applied.

With the development of neural networks, image compression methods based on machine learning and deep learning have merged \cite{toderici2017full,toderici2016variable,balle2016end}. They use neural networks instead of the encoder and decoder to achieve image compression. As for image coding standards, there are some mature instances. JPEG\cite{7924246} and JPEG-2000\cite{1037564} are based on discrete cosine transform \cite{1672377} and wavelet transform \cite{192463}, respectively. JPEG-LS\cite{1999Information} is specially designed for lossless compression, adopting predictive coding, Golomb coding and Run-Length coding.

\subsection{Related Work}

The earliest coding approaches using symbols and locations can be traced back to Symbol-Based coding \cite{1672419}. A picture is denoted as a set of frequently occurring sub-images, called symbols. Storing repeated symbols only once can compress images significantly, especially in document storage where the symbols are usually character bitmaps that are repeated many times. But Symbol-Based coding is hard to generalize to other scenarios owing to the need of redesigning symbols. There are also some methods based on shape coding \cite{4379615,8633164}, but none of them consider both shapes and locations at the same time. Fractal block coding \cite{128028} relies on the assumption that image redundancy can be efficiently exploited through self-transformability on a blockwise basis and it can approximate an original image by a fractal image. However, it is mainly used in lossy compression because it is tough to find a great deal of identical blocks from only one image.
 
 Using database to find similar features is an active research topic in the field of image compression in recent years. In\cite{5995616}, it uses an off-the-shelf image database to find patches that are visually similar to each region of interest of the unknown input image  according to associated local descriptors. These patches are then warped into input image domain according to interest region geometry and seamlessly stitched together. In  \cite{6298446}, it makes use of external image contents to reduce visual redundancy among images by using SIFT descriptors \cite{lowe2004distinctive}. In \cite{6410041}, it proposes a method of cloud-based image coding that no longer compresses images pixel by pixel and instead tries to describe images and reconstruct them from a large-scale image database via the descriptions. In \cite{8241705}, it adopts a semi-local approach to exploit inter-image correlation by using information from external image similarities. In \cite{8272460}, it proposes a cloud storage system that reduces the storage cost of all uploaded JPEG photos, at the expense of a controlled increase in computation mainly during download of requested image subset. In \cite{8024041}, it proposes a novel framework for image set compression based on the rate-distortion optimized sparse coding.

\subsection{Soft Compression}
Soft compression is a lossless image compression method that is based on information theory and statistical distribution to eliminate coding redundancy and spatial redundancy at the same time. It was first proposed in \cite{9247990}, using locations and shapes to represent a binary image. 

The set of shapes used in soft compression is not designed by experts, but searched from dataset. Different datasets may have diverse codebooks for shapes, which ensures the adaptability of soft compression. At the same time, the codebook corresponding to each dataset is also complete, which contains all the possibilities of the smallest shape, which makes soft compression workable. Due to the adaptability and completeness of codebooks of soft compression, it can always achieve lossless compression effect for any image and any codebook. When the codebook and image match well, it will result in a better compression ratio.

The main idea of soft compression is to represent an image by a set of triplets $(x_i,y_i,S_i)$, where $(x_i,y_i)$ denotes the position of shape $S_i$ in an image. The set of shapes is obtained by searching in the training set. After that, the set of codewords and codebook are obtained by variable length coding for the set of shapes according to the size and frequency of each shape. When an image is encoded, it will be transformed into  a set of triplets $(x_i,y_i,C_i)$ according to the codebook, where $C_i$ is the codeword of shape $S_i$. When decoding, we also require to decode the compressed data into a set of triplets $(x_i,y_i,S_i)$ according to the codebook, and then fill shapes in the corresponding locations to reconstruct the original image.

Soft compression is instrumental in reducing storage space and communication bandwidth in the process of transmitting and storing the same kind of images. When two sides communicate, the transmitter only needs to send the compressed data instead of the whole picture to the receiver in the case that both sides have identical codebooks.

In this paper, we try to answer the following two fundamental problems for lossless image compression.

\ding{172} How to detect an image to be compressible in theory? In other words, what is the value of compressible indicator function for an image?

\ding{173} If one image is compressible, how to find a way to compress it through increasing the value of compressible indicator function.

These two problems are great challenges in the theory of image compression.

This paper is organized as follows. We first introduce a new concept, compressible indicator function with regard to image by using information theory and then use it to evaluate the performance of soft compression in Section II. In Section III, some soft compression algorithms for binary image, gray image and multi-component image are proposed. Then we give the experimental results and theoretical analysis in Section IV. Finally, we conclude this paper in Section V.

\section{Theory}
Soft compression is based on statistics and information theory to achieve the image compression effect. Since the image has characteristics of spatial redundancy and coding redundancy, soft compression is committed to eliminating these two kinds of redundancy at the same time by filling an image with locations and shapes. In this section, we introduce the theory of soft compression. 
\subsection{Preliminary}
\subsubsection{Information Theory}
Information theory provides the answer to the lower bound of data compression. For an image, the minimum number of bits per pixel needed is given by formula (\ref{Eq:Ln}), which is the entropy rate of a random process.

\begin{definition}
	Let $Z$ be a discrete random variable with alphabet $\mathcal{Z}$ and probability mass function $p(z)=\text{Pr} \{Z=z\},z \in \mathcal{Z}$. The entropy \cite{shannon1948mathematical} $H(Z)$ of a discrete random variable $Z$ is defined by
	\begin{equation}
	H(Z)=-\sum_{z\in \mathcal{Z}}p(z)\log p(z)
	\end{equation}
\end{definition}
The entropy is a measure of the average uncertainty in the random variable and also clearly shows the required number of bits on average to describe the random variable\cite{cover1999elements}. In this paper, we take all logarithms to base 2 so that entropy is measured in bits unless otherwise specified.

\begin{definition}
	\label{Hp}
	Supposed that Z is a random variable with only two events, i.e.
	\begin{equation}
	Z=\left\{
	\begin{aligned}
	&  0~~~\text{with probability }p \\
	&  1~~~\text{with probability }1-p \\
	\end{aligned}
	\right.
	\end{equation}
	Then the entropy of $Z$ is given by
	\begin{equation}
	H(Z)=-p\log p -(1-p) \log (1-p) {\buildrel \rm def \over = H(p)}
	\end{equation}
\end{definition}
Note that $H(p)$ is a concave function of $p$ and equals 1 when $p=0.5$. In the case where $p=0$ or $p=1$, $H(p)$ reaches its minimum value of 0 because the random variable becomes a constant due to the lack of randomness.
 
If the random variable represented by each pixel in an image is independently and identically distributed, then the minimum expected number of bits required per pixel is the entropy of this random variable. However, for an actual image, the probability distribution of each pixel cannot be independently and identically distributed. Due to the spatial correlation, the minimum expected number of bits required for a pixel is the entropy rate of the random process corresponding to an image. How to evaluate it is still an open problem in the literature.

\subsubsection{Image Fundamentals}
Let $I$ denote a digital image with intensity levels in the range $[0,D-1]$ whose row and column dimensions are $M$ and $N$, respectively. $r_k$ is the $k$-th intensity value and $n_k$ is the number of pixels in the image $I$ with intensity $r_k$\cite{gonzalez2004digital}.
We define $X$ as a discrete random variable with probability mass function $p(x_k)$
\begin{equation}
	 p(x_k)=\text{Pr}\{X=r_k\}={n_k \over MN}~~k=0,1,2,...,D-1
\end{equation}
$X$ reflects the frequency distribution of the intensity value of pixels in an image.

Let $Y$ denote another discrete random variable with $X$ removed from event $r_0$ and $p=\text{Pr}\{X=r_0\}$, then
\begin{equation} 
	p(y_k) = {p(x_k) \over 1-p}~~~k=1,2,...,D-1
\end{equation}
$Y$ reflects the frequency distribution of the remaining intensity values after removing $r_0$ from $X$ in an image.

\begin{lemma}
	\label{Lemma:1}
	Let $H(X)$ and $H(Y)$ denote the entropy of $X$ and $Y$, respectively. then,
	\begin{equation}
		H(Y) ={H(X)-H(p) \over 1-p}
	\end{equation}
\end{lemma}
where $H(p)$ comes from Definition \ref{Hp}.

\noindent
\textbf{Proof.}
\begin{align}
	H(Y) &= -\sum_{k=1}^{D-1}p(y_k)\log p(y_k)  \\
	     &= -\sum_{k=1}^{D-1}p(y_k)\log {p(x_k) \over 1-p} \\
	     &= -\sum_{k=1}^{D-1}p(y_k)\log {p(x_k)} + \sum_{k=1}^{D-1}p(y_k)\log (1-p)  \\
	     &= -\sum_{k=1}^{D-1}{p(x_k) \over 1-p}\log p(x_k) + \log (1-p) \\
	     &= {1 \over 1-p}[-\sum_{k=1}^{D-1}p(x_k)\log p(x_k)] + \log (1-p) \\
	     &= {1 \over 1-p}[H(X)+p \log p] + \log (1-p) \\
	     &= {H(X)+p\log p+(1-p)\log (1-p) \over 1-p} \\
	     &= {H(X)-H(p) \over 1-p}   
\end{align}
$\hfill\blacksquare$ 

\subsection{Soft Compression}
Soft compression is a lossless image compression method which aims to use locations and shapes to denote an image. The purpose of soft compression is to find essential shapes of images.

We define $\mathcal{S}$ and $\mathcal{C}$ as the finite set of shapes and codewords in soft compression, respectively. Let $T$ denote the total number of operations in the process of soft compression for an image $I$\cite{9247990}.

Let $S_i \in \mathcal{S},i=1,...,T$ be the shape used in the $i$-th operation, $C_i \in \mathcal{C},i=1,...,T$ be the codeword of $S_i$, the location of shape $S_i$ in an image is defined as a pair $(x_i,y_i)$. We use $F_i(S_i),i=1,...,T$  to denote filling the image with shape $S_i$ at position $(x_i,y_i)$ in the $i$-th operation. 

Then, soft compression can be formulated as the following optimization problem:
\begin{equation}
\centering
\begin{split}
\min~&\sum_{i=1}^T[l(C_i)+l(x_i,y_i)]\\
&\text{s.t.}~I=\sum_{i=1}^TF_i(S_i)
\label{Eq:optimization2}
\end{split}
\end{equation}
where $l(C_i)$ is the number of bits in the codeword $C_i$ and $l(x_i,y_i)$ is the length of bits needed to represent a location pair $(x_i,y_i)$. The constraint reflects that the original image $I$ can be reconstructed by filling shapes in the corresponding positions through $T$ operations. That is, soft compression is lossless. The core goal of soft compression algorithm design is to find $\mathcal{S}$ and $\mathcal{C}$, so as to encode images efficiently.

\subsection{Image Compression}
\begin{definition}
	We define compressible indicator function (CIF) with respect to $p$ as
	\begin{equation}
	C(p) = {H(p) \over 1-p}~~~~p \in[0,1]
	\end{equation}
	For an image whose probability of $r_0$ is $p$, we define compressible indicator value (CIV) as the value of CIF.
\end{definition}
Compressible indicator function $C(p)$ is derived from Lemma \ref{Lemma:1}, which represents the sparsity of an image. The larger it is, the more sparse the image is, and the image has more capacity to be compressed. The basic properties of compressible indicator function can be summarized as follows.

\begin{theorem}
	\label{Theorem:CC}
	
	\ding{172} $C(p) \geq 0$
	
	\ding{173} $C(p)$ is monotonically increasing.
\end{theorem}

\noindent
\textbf{Proof.}

\ding{172} $0\leq p \leq 1$ implies that $H(p) \geq 0$, which can reach the conclusion that $C(p) \geq 0$.

\ding{173} The derivative of $C(p)$ with respect to $p$ is
\begin{equation}
C'(p)=-{\log p \over (1-p)^2} \geq 0
\end{equation}
therefore, $C(p)$ is monotonically increasing. $\hfill\blacksquare$ 
\\

\begin{definition}
	we use $B_{nb}$ to denote the number of bits needed to encode an image with natural binary code, $B_{hf}$ for Huffman coding, and $B_{sc}^n$ for soft compression where the size of shapes is ranging from 1 to $n$, i.e. soft compression with $n$ order. $B_{hf,min}$ and $B_{sc,min}^n$ represent their minimum values, respectively.
\end{definition}
Let $L_{hf}$ denote the average number of bits required to represent each pixel with Huffman coding, and $L_{sc}^1$ for soft compression where the size of all shapes is one. Then,
\begin{align}
& B_{nb} = MN \log D \\
& B_{hf} = MNL_{hf} \\ 
& B_{sc}^1 = MN(1-p)(L_{sc}^1+L_{W})  \\
& B_{hf,min} = MNH(X) \\
& B_{sc,min}^1 = MN(1-p)(H(Y)+L_{W})
\end{align}
where $L_{W}$ is the average number of bits required to represent a location with soft compression.

\begin{theorem}
	\label{theorem:C(p)}
	If $C(p) \geq L_W$ and $H(X) > 0$, the minimum number of bits needed with 1 order soft compression is less than that with Huffman coding, i.e. $B_{sc,min}^1 \leq B_{hf,min}$, and the relative compression ratio $R'$ is
	\begin{equation}
	R' = 1 + (1-p){ {C(p)-L_W}\over H(X)}
	\end{equation}

\end{theorem}

\noindent
\textbf{Proof.}
\begin{align}
B_{sc,min}^1&=MN(1-p)(H(Y)+L_{W}) \\
&=MN(1-p)[{H(X)-H(p) \over 1-p}+L_{W}] \label{EqIn:1} \\
&=MN[H(X)-H(p)+(1-p)L_{W}]\\
&=B_{hf,min} + MN[(1-p)L_{W}-H(p)]
\end{align}
Equation (\ref{EqIn:1}) uses Lemma \ref{Lemma:1}.

To get the result $B_{sc,min}^1 \leq B_{hf,min}$, we can reach the conclusion that
\begin{equation}
{H(p) \over 1-p} = C(p) \geq L_{W}
\end{equation}
From Theorem \ref{Theorem:CC}, we know that $C(p)$ increases monotonically in $(0,1)$.
Due to the non-negativity of entropy and the trivial case for $H(X)=0$, we mainly consider the case where $H(X)>0$, then 
\begin{align}
R' &= {B_{hf,min} \over B_{sc,min}^1} \\
   &= {{B_{sc,min}^1 - MN[(1-p)L_W-H(p)]} \over {B_{sc,min}^1}} \\
   &= 1- {MN[(1-p)L_W-H(p)] \over B_{sc,min}^1} \\
   &= 1 - {{MN[(1-p)L_W-H(p)]} \over {MNH(X)}} \\
   &= 1 + { H(p)-(1-p)L_W \over H(X)} \\
   &= 1 + (1-p){ {C(p)-L_W}\over H(X)}
\end{align}
It completes the proof.$\hfill\blacksquare$ 
\\

Theorem \ref{theorem:C(p)} gives a threshold about the average number of bits required to represent a location with soft compression by using CIF. When $L_W$ is less than this threshold, the minimum number of bits needed to represent an image with soft compression is lower than that of Huffman coding.

In general, we use the minimum value to approximately replace the actual value needed for compression, which is convenient for theoretical analysis. Theorem \ref{theorem:C(p)} indicates that for an image whose compressible indicator value is greater than the average number of bits required for a location, soft compression is better than Huffman coding in terms of compression ratio. It also points out that the larger the compressible indicator value, the higher the compression ratio.

\begin{figure}
	\centering{\includegraphics[width=\columnwidth]{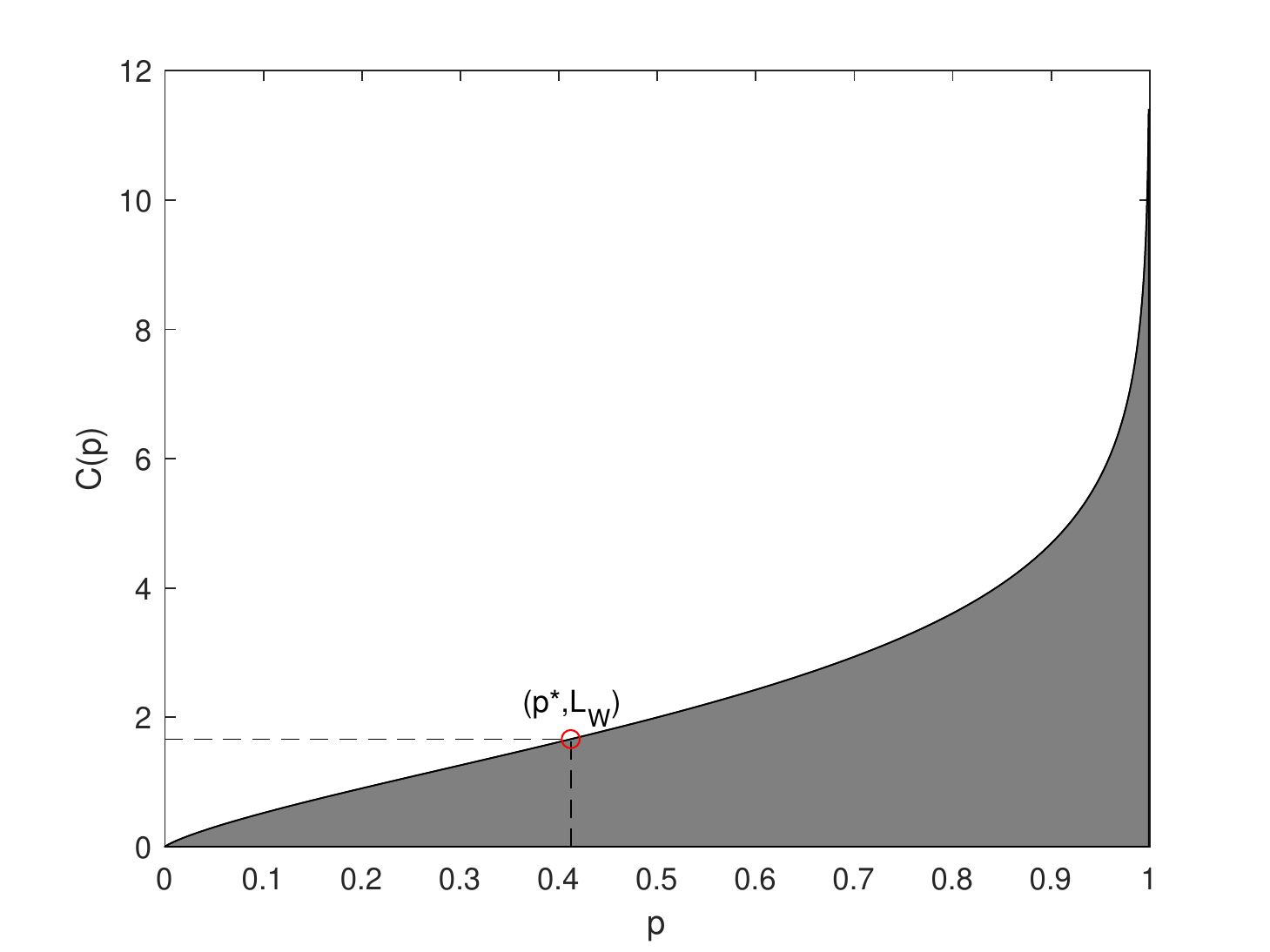}}
	\caption{Compressible indicator function versus p.}
	\label{Fig:CompressionIndicatorFunction}
\end{figure}

Fig. \ref{Fig:CompressionIndicatorFunction} illustrates the relationship between compressible indicator function $C(p) $and $p$. Theorem \ref{theorem:C(p)} presents a rule to use soft compression. Given an image, we should first evaluate the value of compressible indicator function according to $p$, and then judge whether it is suitable for soft compression. That is to say, if $(p, L_W)$ is in the gray area, the minimum number of bits needed with 1 order soft compression is less than that with Huffman coding. As shown in Fig. \ref{Fig:CompressionIndicatorFunction}, it gives an answer to the first fundamental problem of lossless image compression.

\begin{lemma}
	\label{Lemma:entropybound}
	Let $Y_n$ represent the frequency distribution of a shape whose size is $n$ in an image, then
	\begin{equation}
	H(Y_n) \leq nH(Y)
	\end{equation}
\end{lemma}
\noindent
\textbf{Proof.} From the independence bound on entropy, one can come to this conclusion.$\hfill\blacksquare$ 
\\

\begin{theorem}
	\label{theorem:n1}
	If $L_W$ is a constant for different orders of soft compression, then $B_{sc,min}^n \leq B_{sc,min}^1$.
\end{theorem}

\noindent
\textbf{Proof.} We use $N_1$ to represent the number of shapes with size 1 used in an image, and $N_n$ is the number of shapes with size $n$. The derivation can be seen from (\ref{Eq:total-1-1}) to (\ref{Eq:total-1-2}). $\hfill\blacksquare$ 
\\

\begin{figure*}
\begin{align}
B_{sc,min}^n &= N_1[H(Y)+L_W] + N_2[H(Y_2)+L_W]+...+N_n[H(Y_n)+L_W]  \label{Eq:total-1-1} \\
             &\leq [N_1+2N2+...+nN_n]H(Y)+[N_1+N_2+...+N_n]L_W \\
             &\leq [N_1+2N_2+...+nN_n][H(Y)+L_W] \\
             &= MN(1-p)[H(Y)+L_W] \\
             &= B_{sc,min}^1 \label{Eq:total-1-2}
\end{align}
\end{figure*}

Theorem \ref{theorem:C(p)} and \ref{theorem:n1} inspire us that in image compression, we can improve the compression ratio from two aspects: one is to increase the compressible indicator value of an image; the other is to reduce the number of bits needed to represent locations. On one hand, compressible indicator value can  be increased by predictive coding which only encodes prediction error. On the other hand, Golomb coding maybe a promising method for encoding the distance between each location and the previous location in order to reduce $L_W$. 
\section{Implementation Algorithm}
In this section, we try to answer the second problem, to find some ways to increase the compression ratio of images by raising compressible indicator value. We first introduce the algorithms of soft compression for different image formats, including binary image, gray image and multi-component image. In fact, three different algorithms vary in specific steps but all try to make use of locations and shapes to represent an image from the perspective of information theory and statistical distribution.

\subsection{Binary Image}
Binary image is quite suitable for lossless compression using soft compression method, because the pixel of binary image has only two intensity values. The probability of occurrence of $r_0$ can always be greater than or equal to 0.5 (through reverse operation), which ensures that the compression indicator value is greater than or equal to 2.

The soft compression algorithm for binary image was proposed in \cite{9247990}, which has an excellent compression effect. Although \cite{9247990} introduced the algorithm for binary image, it didn't give a  theoretical analysis. We will analyze the experimental results by using compressible indicator function in Section IV.
\subsection{Gray Image}
An important procedure of soft compression algorithm for gray image is to divide the image into two layers, shape layer and detail layer. In fact, compressible indicator value of shape layer is usually large, so the combinations of locations and shapes are used to encode. While compressible indicator value of detail layer is relatively small, and other common coding methods can be used in detail layer. Soft compression algorithm for gray image includes predictive coding, negative-to-positive mapping, layer separation, shape search, codebook generation and so on. We will introduce the vital steps one by one.
\subsubsection{Predictive Coding and Negative-to-Positive Mapping}
Predictive coding \cite{boon2000image} is a way to transform spatial redundancy into coding redundancy by means of prediction. The main idea of predictive coding is to calculate the prediction value of the pixel according to the spatial correlation, so as to encode the prediction error.

\begin{figure}
	\centering{\includegraphics[width=2in]{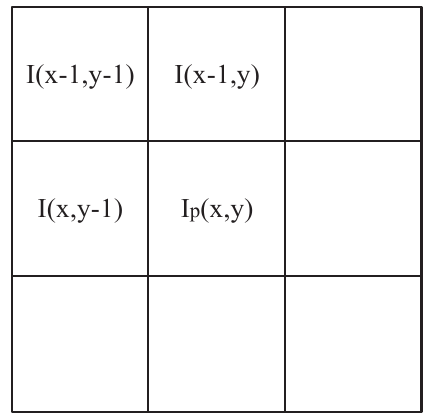}}
	\caption{The spatial correlation between pixels with predictive coding.}
	\label{Fig:Prediction}
\end{figure}

Let $I(x,y)$,$I_P(x,y)$ and $I_E(x,y)$ denote the pixel intensity value, predictive value, prediction error at location $(x,y)$, respectively. The image predictive coding method \cite{1999Information} we use is shown in the formula (\ref{Eq:prediction}), where the spatial relationship between pixels is displayed in Fig. \ref{Fig:Prediction}.
\begin{figure*}
	\begin{equation}
	\label{Eq:prediction}
	I_P(x,y)=\\
	\left\{
	\begin{aligned}
	&  \min(I(x,y-1),I(x-1,y))~~&&\text{if }I(x-1,y-1) \geq \max(I(x,y-1),I(x-1,y)) \\
	&  \max(I(x,y-1),I(x-1,y))~~&&\text{if }I(x-1,y-1) \leq \min(I(x,y-1),I(x-1,y)) \\
	&  I(x,y-1)+I(x-1,y)-I(x-1,y-1)~~~&&\text{others} \\
	\end{aligned}
	\right.
	\end{equation}
\end{figure*}

After getting the predictive value, one can get the prediction error by formula (\ref{Eq:Prediction error}). The range of the prediction error is in $[-D+1,D-1]$, which is different from the range of pixel intensity value $[0,D-1]$. We use formula (\ref{Eq:Mapping}) to map prediction error from negative to positive, which is conducive to the subsequent layer separation.

\begin{equation}
\label{Eq:Prediction error}
I_E(x,y)=I(x,y)-I_P(x,y)
\end{equation}

\begin{equation}
\label{Eq:Mapping}
I'(x,y)=\left\{
\begin{aligned}
&  2I_E(x,y)   && I_E(x,y) \geq 0 \\
&  -2I_E(x,y)-1 && I_E(x,y) < 0 \\
\end{aligned}
\right.
\end{equation}
After predictive coding and mapping, the proportion of pixel value $r_0$ increases, which also leads to the increase of compressible indicator value. The image will be more conducive to the use of soft compression for image coding. Another reason for adopting predictive coding and mapping is that the same image will produce the same result after these two operations. The reversibility of steps ensures the lossless performance of soft compression algorithm.
\subsubsection{Layer Separation}
Soft compression is suitable for images with large compressible indicator value. After the previous steps, the compressible indicator value of the image is significantly increased. We want to further increase it. On one hand, we observe that the probability of the prediction error decreases with the increase of the error intensity value. On the other hand, through a proper separation, one image with large compressible indicator value and another with small compressible indicator value can be generated.

Layer separation and bit-plane coding \cite{4501787} are similar, but bit-plane coding focuses on decomposing a multilevel image into a series of binary images and compressing each binary image via one of several well-known binary compression methods, which will produce many layers. Layer separation produces only two layers, shape layer $I'_S$ and detail layer $I'_D$. 

$I'$ is divided into shape layer $I'_S$ and detail layer $I'_D$ via formula (\ref{Eq:Layer operation-1}) and (\ref{Eq:Layer operation-2}).
\begin{align}
I'_S(x,y) = I'(x,y)~//~2^l \label{Eq:Layer operation-1}\\
I'_D(x,y) = I'(x,y)~\%~2^l \label{Eq:Layer operation-2}
\end{align}
where $//$ and $\%$ represent quotient and remainder operation, respectively. Layer interface $l$ is a constant between 0 and $\log D$ which can be given in advance by searching or experience.

Because of the compressible indicator value of the shape layer is usually large, the combinations of locations and shapes are used to encode. While compressible indicator value of detail layer is relatively low, other coding methods such as Huffman coding \cite{huffman1952a} and block coding can be used to compress.

\subsubsection{Shape Search and Codebook Generation}
The shape layer uses locations and shapes to encode, the set of shapes directly determines the compression ratio and coding efficiency of an image, so how to find the set of shapes is vital.

$A$ is an $M \times N$ matrix whose components are in $[0,2D-2]$, $u_i$ and $v_j$ are vectors, representing the $i$ row and the $j$ column of $A$, respectively. The matrix whose $\boldsymbol{u_i}$ and $\boldsymbol{v_j}$ that follows (\ref{Eq:Design-1}) and (\ref{Eq:Design-2}) is compatible for designing shapes.
\begin{align}
&||\boldsymbol{u_i}||_0 \geq {N \over 2}~~~~\forall~1 \leq i \leq M  \label{Eq:Design-1}\\
&||\boldsymbol{v_j}||_0 \geq {M \over 2}~~~~\forall~1 \leq j \leq N \label{Eq:Design-2}
\end{align}
(\ref{Eq:Design-1}) and (\ref{Eq:Design-2}) show that the number of non-zero elements in all rows and columns of a matrix must be no less than half of the row size and column size, respectively. By keeping only the non-zero value blocks in the matrix and removing the zero value blocks, we can get the final shape. Using this method, we can avoid the situation that different matrices produce the same shape.

\begin{figure}
	\centering{\includegraphics[width=\columnwidth]{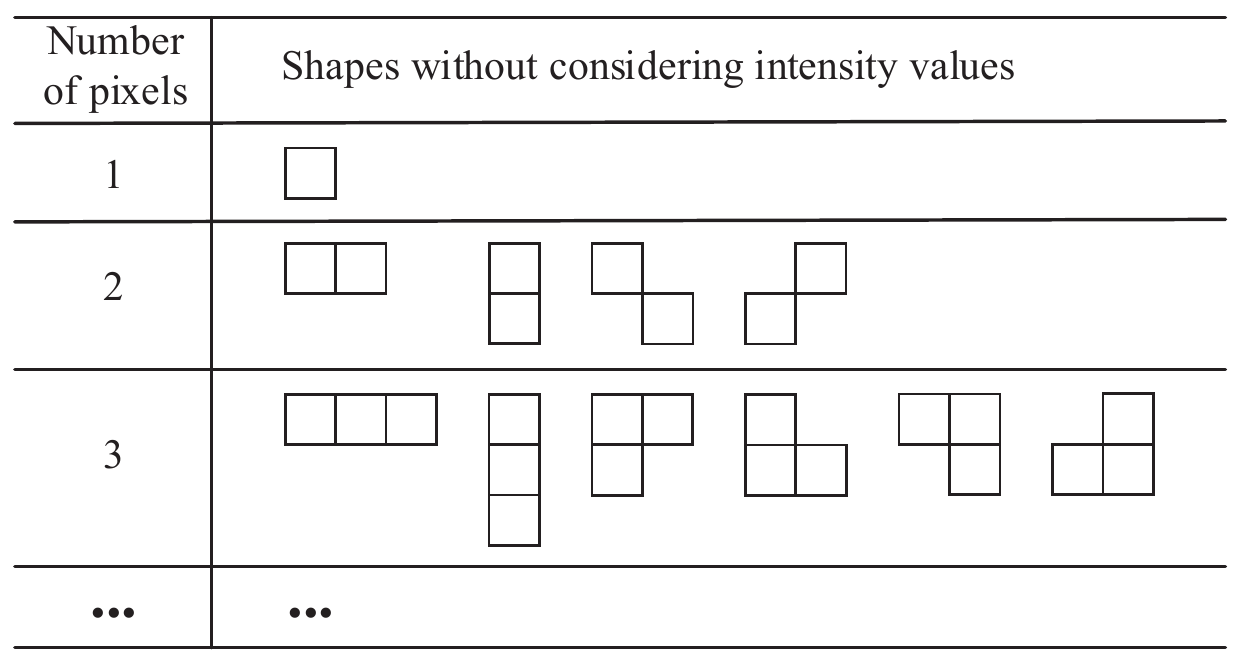}}
	\caption{Shapes generated without considering intensity values according to the criteria (classified by the number of pixels). These are not all shapes in the set but only a part of it.}
	\label{Fig:Shape}
\end{figure}

As illustrated in Fig. \ref{Fig:Shape}, there is a part of shapes. These shapes are classified by size and not considered intensity values. Combine them with error intensity value $[1,2D-2]$ to produce shapes that are actually in use. 

However, not all shapes will appear in the final shape set. In fact, only a small number of shapes can be retained all the time. This is because in the process of searching, the set of shapes will be updated dynamically, and shapes with less frequency will be deleted, so as to ensure that the total number of shapes will not be too much.

After getting the set of shapes, codebook for shape layer can be generated according to the frequency and size of each shape. When searching for the set of shapes used in the shape layer, we also count the frequency distribution of pixel intensity value in the detail layer, so as to generate the codebook for detail layer. In the process of communication, when both sides have the same codebook, the transmitter can directly send compressed data instead of the whole image to the receiver, which is able to greatly reduce the communication bandwidth and storage space.

\subsubsection{Golomb Coding for Locations}
When the shape layer is encoded, a series of locations are generated. By using Golomb coding for the distance difference between each location and the previous location, the number of bits needed to represent these locations can be reduced.

\begin{figure}
	\centering{\includegraphics[width=\columnwidth]{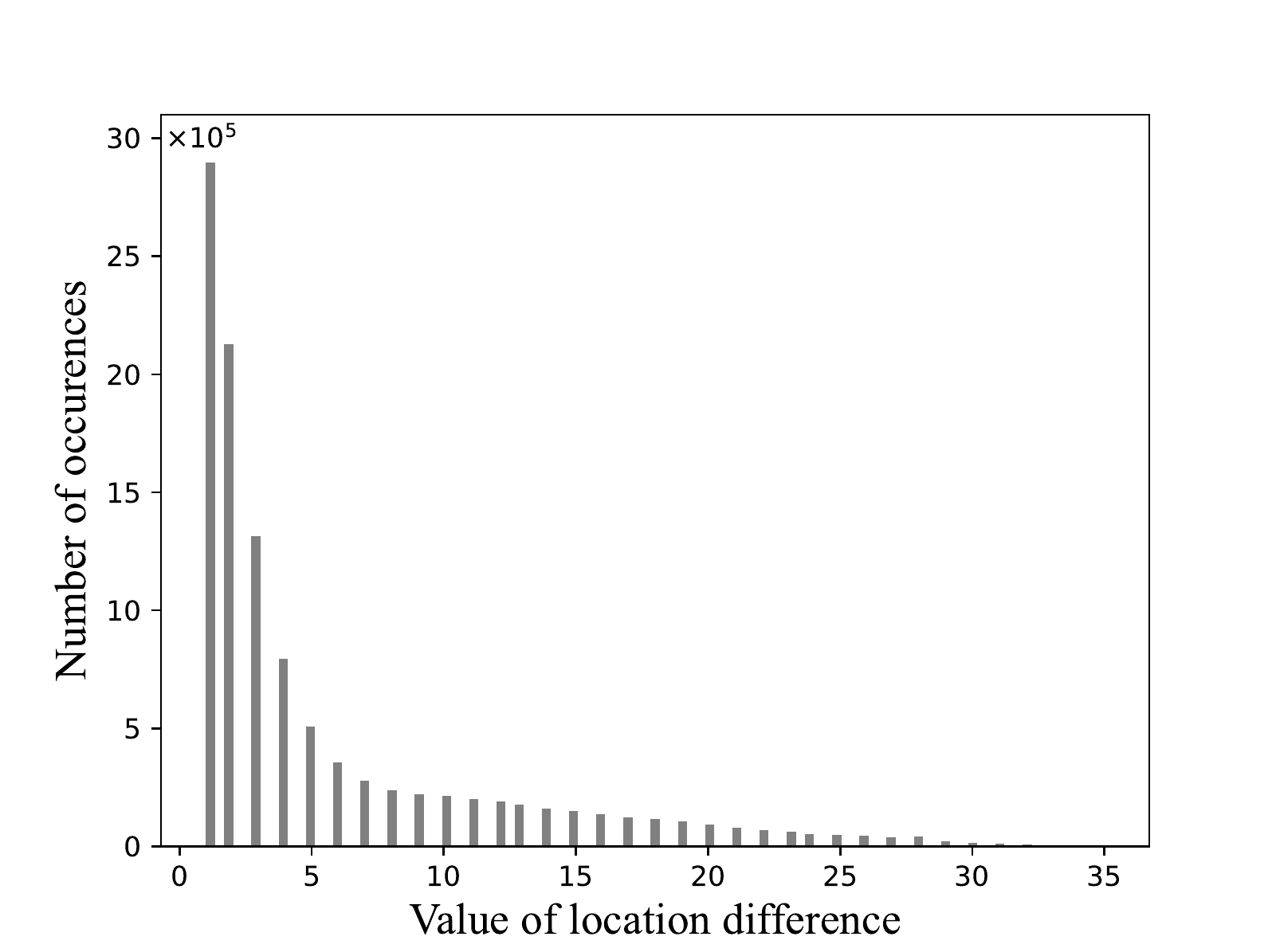}}
	\caption{The frequency distribution of location difference on Fashion-mnist dataset by using soft compression algorithm for gray image.}
	\label{Fig:FreOfLocation}
\end{figure}

Golomb coding \cite{golomb1966run} was designed for non-negative integer input with geometric probability distribution. Golomb coding of one location we used adopts the following steps.
\begin{itemize}
	\item \textbf{Step 1.} Calculate the distance difference $\Delta$ from the previous location.
	\item \textbf{Step 2.} Get a positive integer $m$ by giving or searching in advance.
	\item \textbf{Step 3.} Form the unary code of quotient $\lfloor \Delta/m \rfloor$. (The unary code of an integer $q$ is defined as $q$ 1s followed by a 0.)
	\item \textbf{Step 4.} Let $k = \lceil \log_2m\rceil$, $c=2^k-m$, $r=\Delta \text{ mod } m$, and compute truncated remainder $r'$ such that
	\begin{equation}
	r'=\left\{
	\begin{aligned}
	&  r \text{ truncated to $k-1$ bits}~~&&0 \leq r <c \\
	&  r+c \text{ truncated to $k$ bits}~~   &&\text{otherwise} \\
	\end{aligned}
	\right.
	\end{equation}  
	\item \textbf{Step 5.} Concatenate the results of steps 3 and 4.
\end{itemize}
The location difference obtained by soft compression approximately obeys geometric probability distribution. Fig. \ref{Fig:FreOfLocation} shows the empirical frequency distribution of location difference on Fashion-mnist dataset by using soft compression algorithm for gray image which has a good match to the scenarios of Golomb coding. Under the prior information, using Golomb coding will greatly reduce the space for storing location differences.

\subsubsection{Encoder}
\begin{figure*}
	\centering{\includegraphics[width=7.16in]{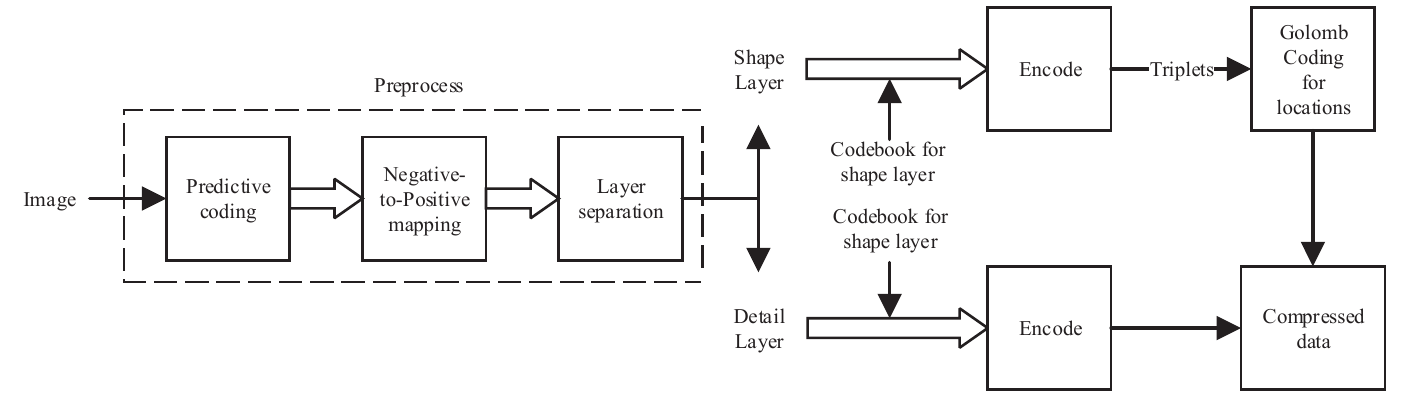}}
	\caption{The overall procedure of Encoder. It uses two different codebooks to encode the shape layer and the detail layer, respectively.}
	\label{Fig:Encoder}
\end{figure*}

When an image is encoded, preprocessing is needed first, which includes predictive coding, negative-to-positive mapping and layer separation. Secondly, two codebooks are used to encode the shape layer and the detail layer, respectively. Thirdly, locations in the shape layer are encoded by Golomb coding. Finally, connect the encoded results of the two layers to obtain the compressed data. Fig. \ref{Fig:Encoder} shows the entire process of Encoder.

\begin{figure}
	\centering{\includegraphics[width=\columnwidth]{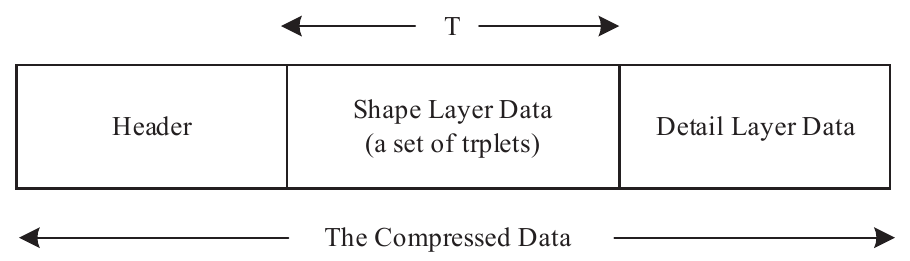}}
	\caption{The composition of compressed data.}
	\label{Fig:Frame}
\end{figure}

Fig. \ref{Fig:Frame} illustrates the composition of the compressed data. Header part contains information about the height and width of an image and the layer interface. Shape layer data and detail layer data carry the encoding results of these two layers, respectively.

\subsubsection{Overall Procedure}

\begin{figure*}
	\centering{\includegraphics[width=7.16in]{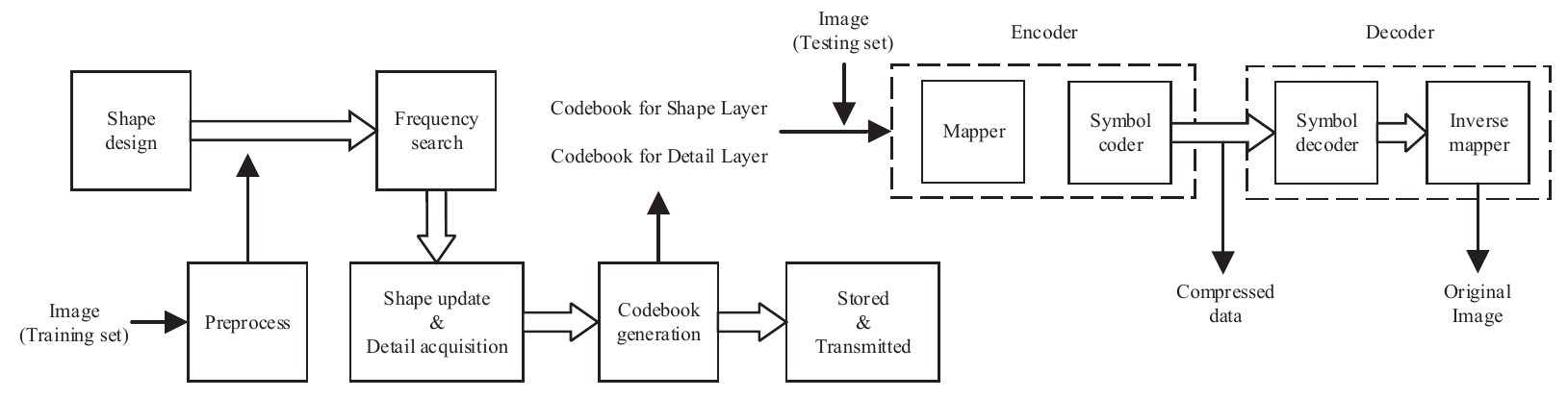}}
	\caption{The overall procedure of soft compression algorithm for gray image including training and testing. The shape design, frequency search, shape update, detail acquisition, codebook generation, and saved codebook are the training stage. On the other hand, the encoder and decoder are the testing stage.}
	\label{Fig:Flowchart}
\end{figure*}

Fig. \ref{Fig:Flowchart} summarizes the overall procedure of soft compression algorithm for gray image. We design the set of shapes firstly, and then search the frequency of each designed shape in the training set while updating the set of shapes by deleting shapes whose frequency are below the reference value. After getting the final set of shapes and its frequency, we use Huffman coding to generate the codebook for the shape layer. In the training stage, we will also acquire the codebook for detail layer at the same time. Codebooks are used in the Encoder and Decoder, which should also be stored or transmitted for subsequent usage.

In the testing stage, the image will be compressed through the Encoder. When the sender wants to communicate with the receiver, it will firstly transmit two codebooks. After both sides of the communication have the same codebook, the transmitted content will be the compressed data instead of the original image.

After storage or transmission, the receiver will get the compressed data. Decoder adopts the opposite structure to the Encoder, and the original image can be reconstructed by inputting the compressed data into Decoder. 

Firstly, the shape $S_i$ is obtained according to the codebook for shape layer, and then the shape $S_i$ is filled in the location $(x_i,y_i)$, shape layer $I'_S$ is acquired by repeating $T$ operations. Secondly, the detail layer $I'_D$ is decoded by the codebook for detail layer from the compressed data. Finally, the original image can be recovered by merging $I'_S$ and $I'_D$, positive-to-negative mapping and anti-predictive coding. Due to the completeness of codebooks, the recovered image is exactly the same as the original image, which ensures the lossless compression.

\subsection{Multi-component Image}
Considering multi-component image, soft compression algorithm for gray image can be used for each component. In this case, soft compression algorithm for multi-component image is equivalent to the combination of several soft compression algorithms for gray image, and the compressed data are also a combination of several components.

\section{Experimental Results and Theoretical Analysis}
In this section, we reveal the experimental results and theoretical analysis of soft compression algorithms.
\begin{definition}
	Supposed that $b$ and $b'$ represent the number of bits required to express the same image by using natural binary code and other coding methods, respectively. The compression ratio $R$ is defined as
	\begin{equation}
	R={b \over b'}
	\end{equation}
	and $R_{avg}$ is defined as the average compression ratio of a class of images.
\end{definition}
The compression ratio reflects the effect of different coding methods. We use it as an significant criterion to measure coding methods in this section.
\subsection{Binary Image}
Soft compression algorithm for binary image \cite{9247990} is tested on the MNIST \cite{lecun1998gradient} dataset and has excellent results. In this subsection, we analyze the experimental results theoretically.

\begin{figure*}
	\centering{\includegraphics[width=7.16in]{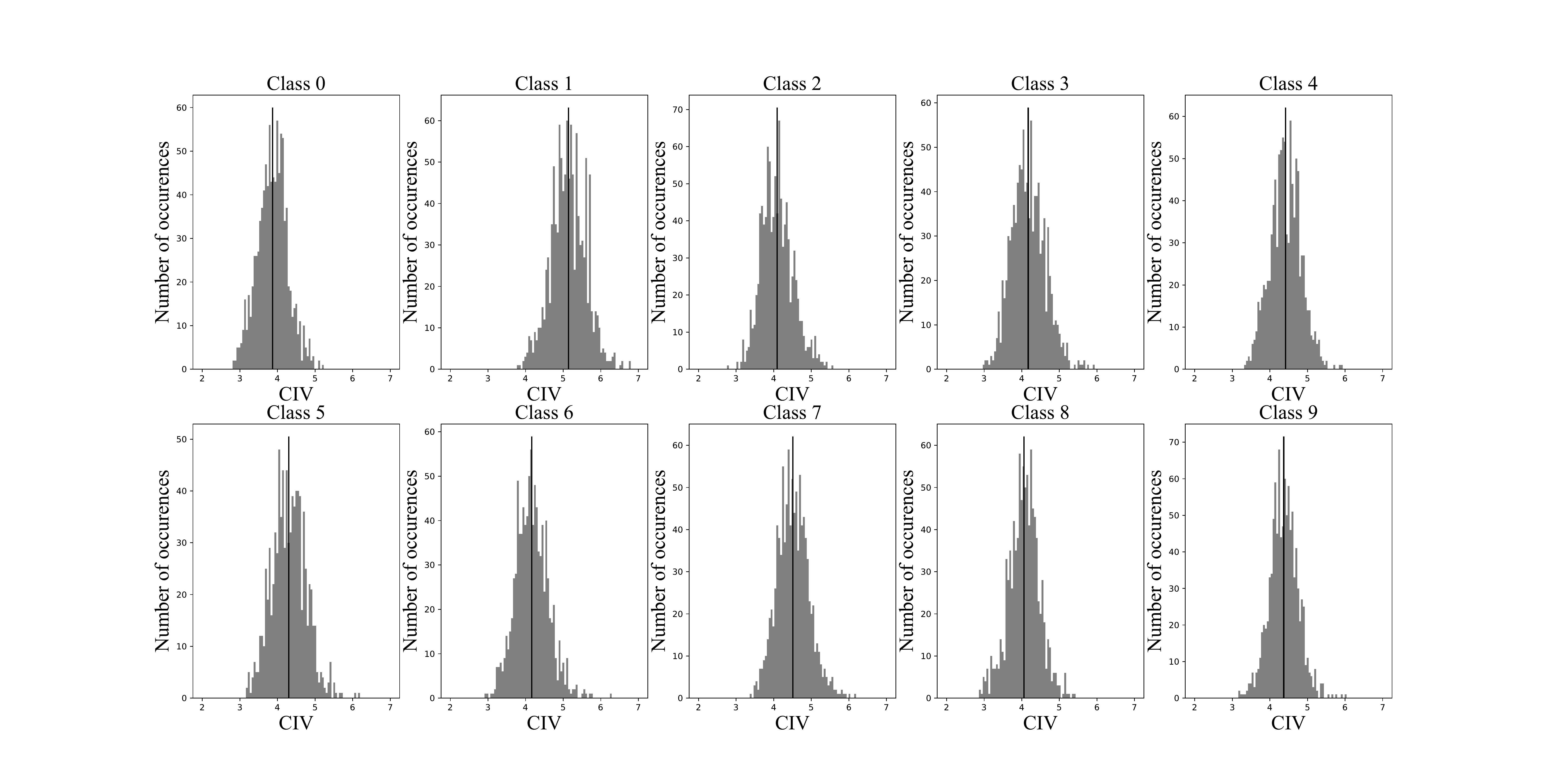}}
	\caption{Average compressible indicator value of different classes on MNIST dataset.}
	\label{Fig:Histogram}
\end{figure*}

MNIST has ten categories, different classes have different compressible indicator values, and the frequency histogram of compressible indicator value is shown in Fig. \ref{Fig:Histogram}. Although they are of different classes, the compressible indicator value is generally subject to the normal distribution.

Table \ref{Tab:Compression indicator function and ratio} illustrates the average compressible indicator
value and compression ratio of MNIST dataset by using the same codebook. From Table \ref{Tab:Compression indicator function and ratio}, it is shown that they are positively related, the larger the average compressible indicator value, the greater the average compression ratio, which is consistent with the theoretical result in Theorem \ref{theorem:C(p)}. It enlightens us that soft compression is suitable for compressing images with large compressible indicator value. Although the compression ratio is not only related to this factor, the compressible indicator value becomes a key element affecting the compression ratio.
\begin{table*}
	\caption{Average compressible indicator value and compression ratio of MNIST dataset by using the same codebook}
	\small
	\setlength{\tabcolsep}{6pt}
	\begin{center}
		\begin{tabular}{ccccccccccc}
			\toprule
			Class& 0    & 1    & 2    & 3    & 4  &5 &6 &7 &8 &9       \\
			\midrule
			CIV      & 3.87 & 5.14 & 4.09 & 4.17 & 4.42 & 4.30 & 4.17 & 4.51 & 4.06 & 4.37   \\
			Compression rate  & 2.84 & 6.02 & 3.17 & 3.20 & 3.77 & 3.40 & 3.20 & 4.05 & 2.81 & 3.52  \\
			\bottomrule
		\end{tabular}
	\end{center}
	\label{Tab:Compression indicator function and ratio}
\end{table*}
\subsection{Gray Image}
In this subsection, we obtain the compression ratio of Fashion-mnist \cite{xiao2017fashion} by using soft compression algorithm for gray image.

Table \ref{Tab:Compression ratio-1-1} illustrates results of soft compression algorithm for gray image on Fashion-mnist, it is the result of cross validation. The first row represents each category of testing set and the first column denotes codebooks generated by the corresponding class in the training set. The value in $(i,j)$ denotes the average compression ratio $R_{avg}$ of $j$-th class of testing set by using the codebook generated by $i$-th class of training set.

It is observed that values on the diagonal are higher than that of the same column, which suggests that soft compression is related to the training set for generating the codebook. The more suitable codebook is used for encoding one class of images, the higher the corresponding compression ratio will be.

In practice, for the same kind of images, it is better to adopt the corresponding codebook. However, the use of different types of codebook for encoding will not produce disgusting consequences, and will not cause loss of images. In fact, codebooks of soft compression are complete, that is to say, for any codebook and any picture, lossless compression can always be attained in the image dataset. The difference lies in diverse compression ratios. 

From Table \ref{Tab:Compression ratio-1-1}, we can draw a conclusion that the compression ratio is both related to codebooks and images. It is reflected in the image compressible indicator value and the similarity between images and codebooks. The larger the compressible indicator value of image is, the higher the similarity between images and codebooks generated by the training set, then the compression ratio will become higher.
\begin{table}
	\caption{Average compression ratio of Fashion-mnist dataset by using soft compression algorithm for gray image(each class has its own codebook)}
	\small
	\setlength{\tabcolsep}{3.5pt}
	\begin{tabular}{ccccccccccc}
		\toprule
		Class  & 0    & 1    & 2    & 3    & 4    & 5    & 6    & 7    & 8    & 9\\
		\midrule
		0 & \textbf{1.72} & 2.43 & 1.58 & 2.07 & 1.57 & 2.18 & 1.55 & 2.27 & 1.67 & 1.78 \\
		1 & 1.67 & \textbf{2.52} & 1.54 & 2.07 & 1.54 & 2.15 & 1.51 & 2.24 & 1.62 & 1.74 \\
		2 & 1.71 & 2.45 & \textbf{1.58} & 2.06 & 1.58 & 2.16 & 1.55 & 2.25 & 1.66 & 1.77 \\
		3 & 1.70 & 2.50 & 1.56 & \textbf{2.10} & 1.56 & 2.22 & 1.54 & 2.30 & 1.65 & 1.78 \\
		4 & 1.71 & 2.44 & 1.58 & 2.06 & \textbf{1.58} & 2.15 & 1.55 & 2.25 & 1.66 & 1.77 \\
		5 & 1.69 & 2.44 & 1.54 & 2.08 & 1.55 & \textbf{2.29} & 1.52 & 2.35 & 1.66 & 1.80 \\
		6 & 1.71 & 2.44 & 1.58 & 2.07 & 1.58 & 2.17 & \textbf{1.55} & 2.26 & 1.66 & 1.78 \\
		7 & 1.68 & 2.43 & 1.53 & 2.06 & 1.54 & 2.26 & 1.52 & \textbf{2.35} & 1.65 & 1.79 \\
		8 & 1.72 & 2.42 & 1.58 & 2.07 & 1.57 & 2.20 & 1.55 & 2.28 & \textbf{1.68} & 1.79 \\
		9 & 1.70 & 2.40 & 1.56 & 2.05 & 1.57 & 2.19 & 1.54 & 2.28 & 1.66 & \textbf{1.79} \\
		\bottomrule
	\end{tabular}
	\label{Tab:Compression ratio-1-1}
\end{table}

Table \ref{Tab:Compression ratio-2} illustrates the comparison results of different methods and shows that soft compression algorithm for gray image is better than some known lossless methods in terms of compression ratio. For one image, if it has a large compressible indicator value, it is a better way to select soft compression to encode it.
\begin{table*}
	\caption{Average compression ratio of Fashion-mnist dataset by using different methods(all classes use the same codebook)}
	\small
	\setlength{\tabcolsep}{6pt}
	\begin{center}
		\begin{tabular}{ccccccccccc}
			\toprule
			Class& 0    & 1    & 2    & 3    & 4 & 5 & 6    & 7    & 8    & 9       \\
			\midrule
			Soft compression & \textbf{1.69} & \textbf{2.48} & \textbf{1.55} & \textbf{2.08} & \textbf{1.55} & \textbf{2.22} & \textbf{1.53} & \textbf{2.30} & \textbf{1.65} & \textbf{1.77} \\
			Huffman coding   & 1.42 & 2.15 & 1.30 & 1.86 & 1.41 & 2.34 & 1.34 & 2.17 & 1.47 & 1.71   \\
			Predictive-Golomb coding  & 1.37 & 1.58 & 1.35 & 1.43 & 1.33  & 1.36 & 1.32 & 1.43 & 1.34 & 1.35  \\
			\bottomrule
		\end{tabular}
	\end{center}
	\label{Tab:Compression ratio-2}
\end{table*}
\subsection{Multi-component Image}
In this subsection, we show an instance of Fundus Image Registration dataset \cite{hernandez-matas2017fire} by using soft compression algorithm for multi-component image. 

\begin{figure*}
	\centering
	\subfigure[Original image]{
		\includegraphics[width=4cm]{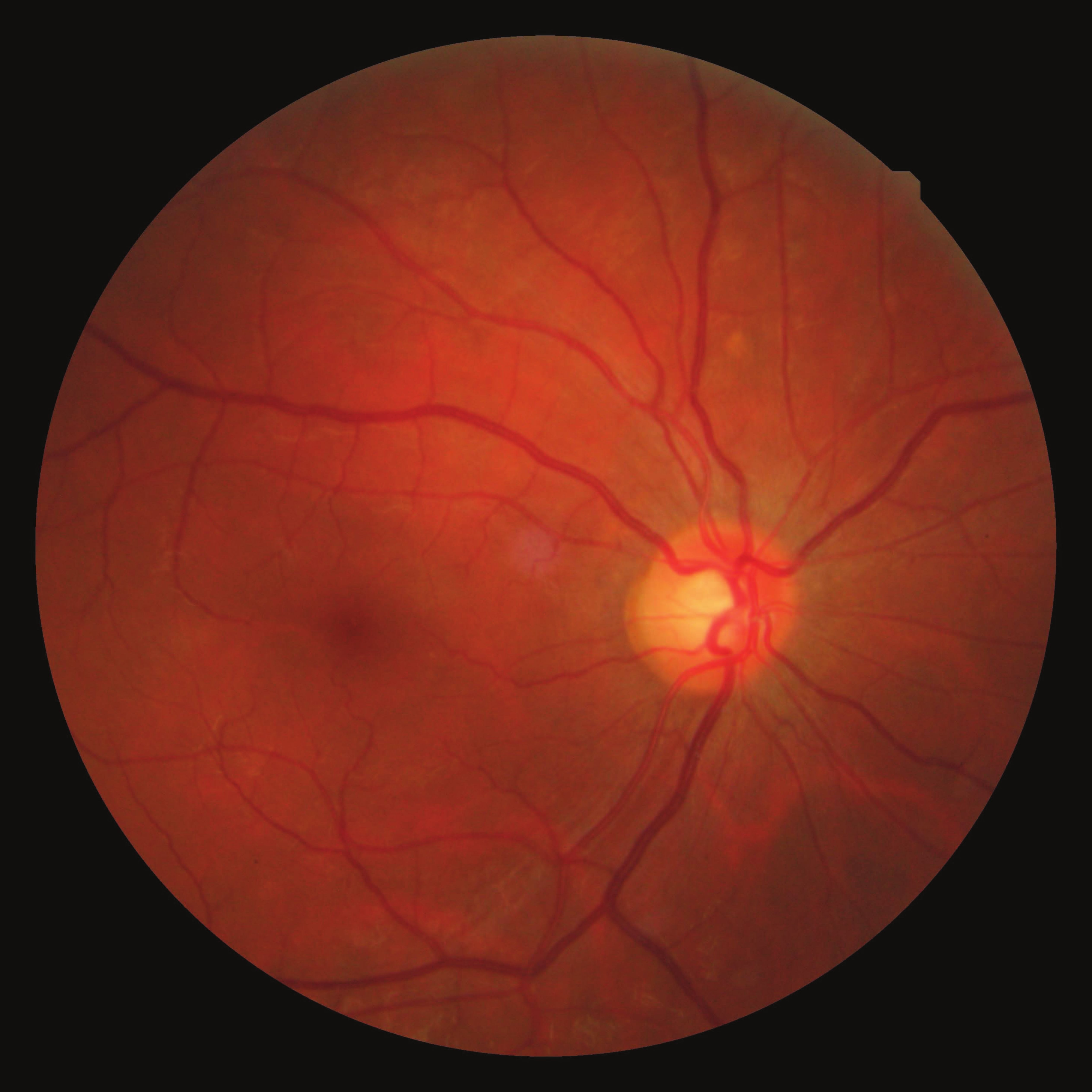}
	}
	\quad
	\subfigure[B-component]{
		\includegraphics[width=4cm]{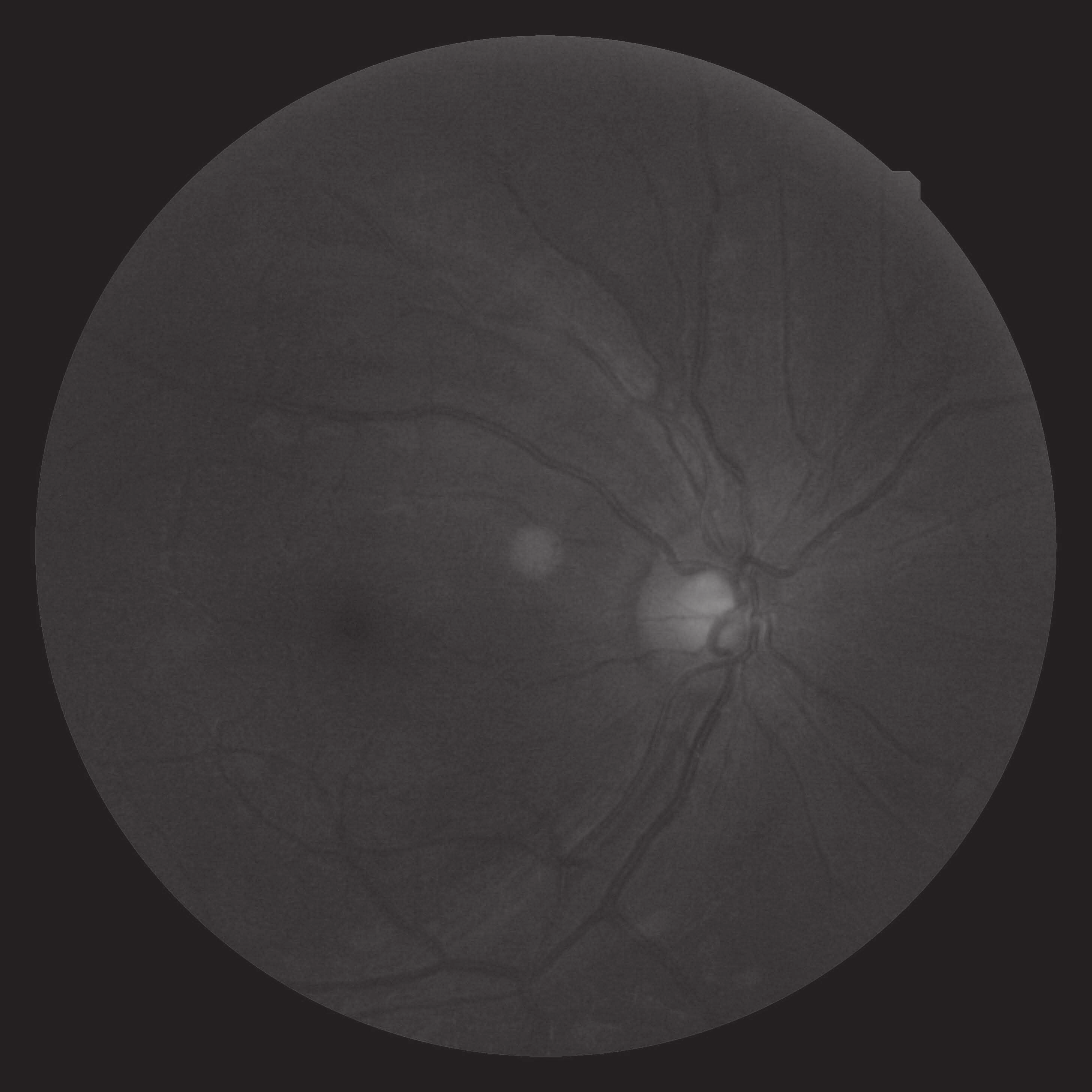}
	}
	\quad
	\subfigure[G-component]{
		\includegraphics[width=4cm]{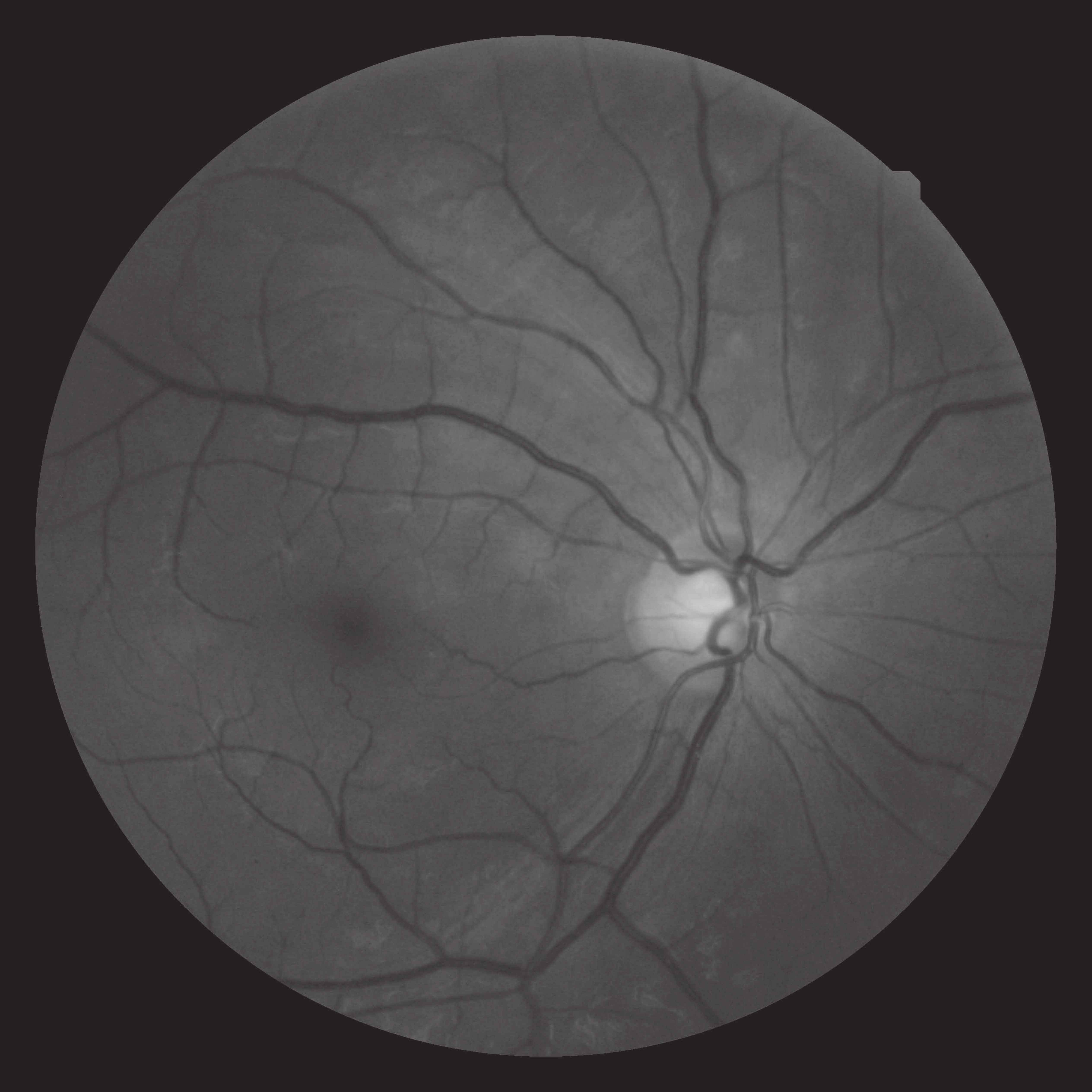}
	}
	\quad
	\subfigure[R-component]{
		\includegraphics[width=4cm]{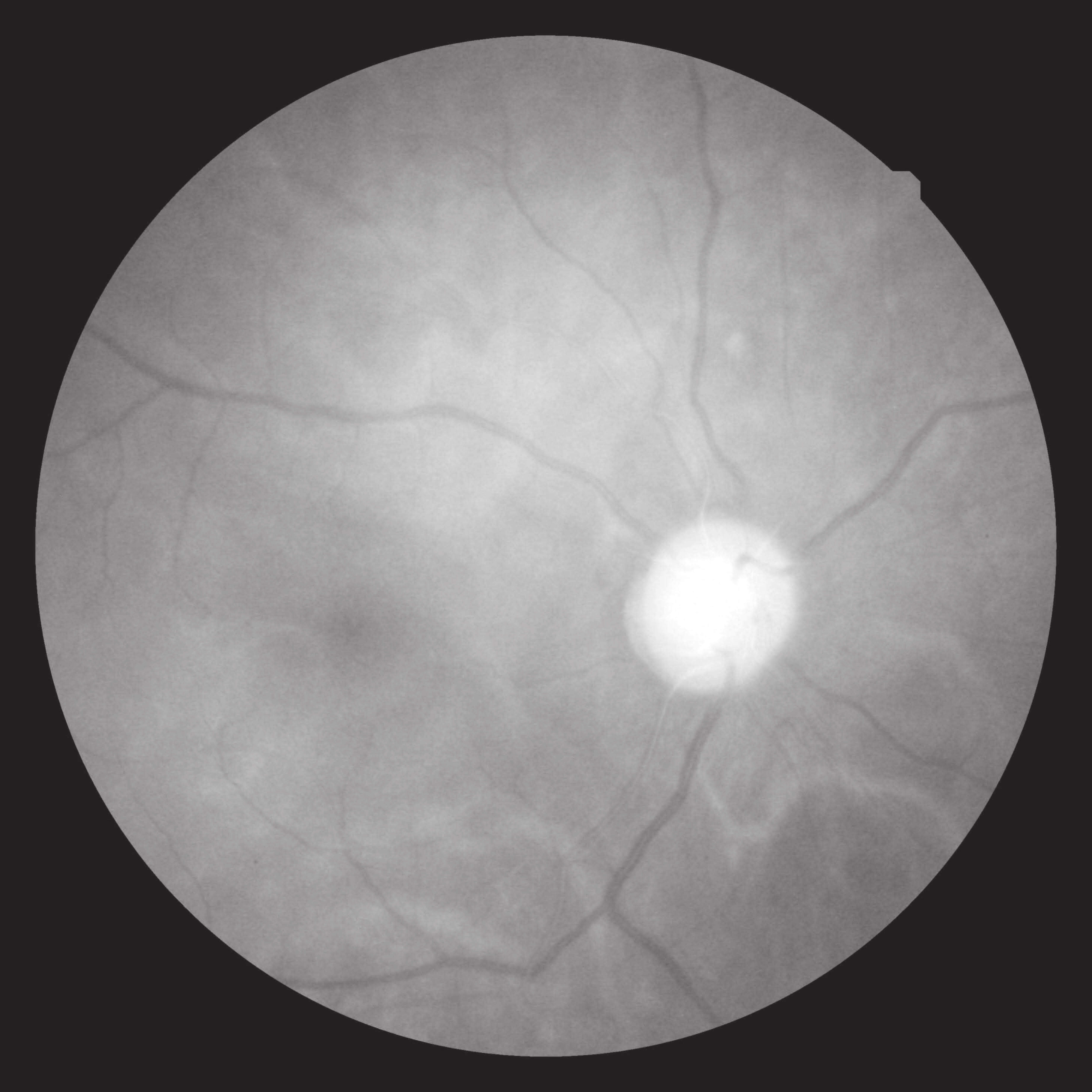}
	}
	\quad
	\subfigure[B-component's shape layer]{
		\includegraphics[width=4cm]{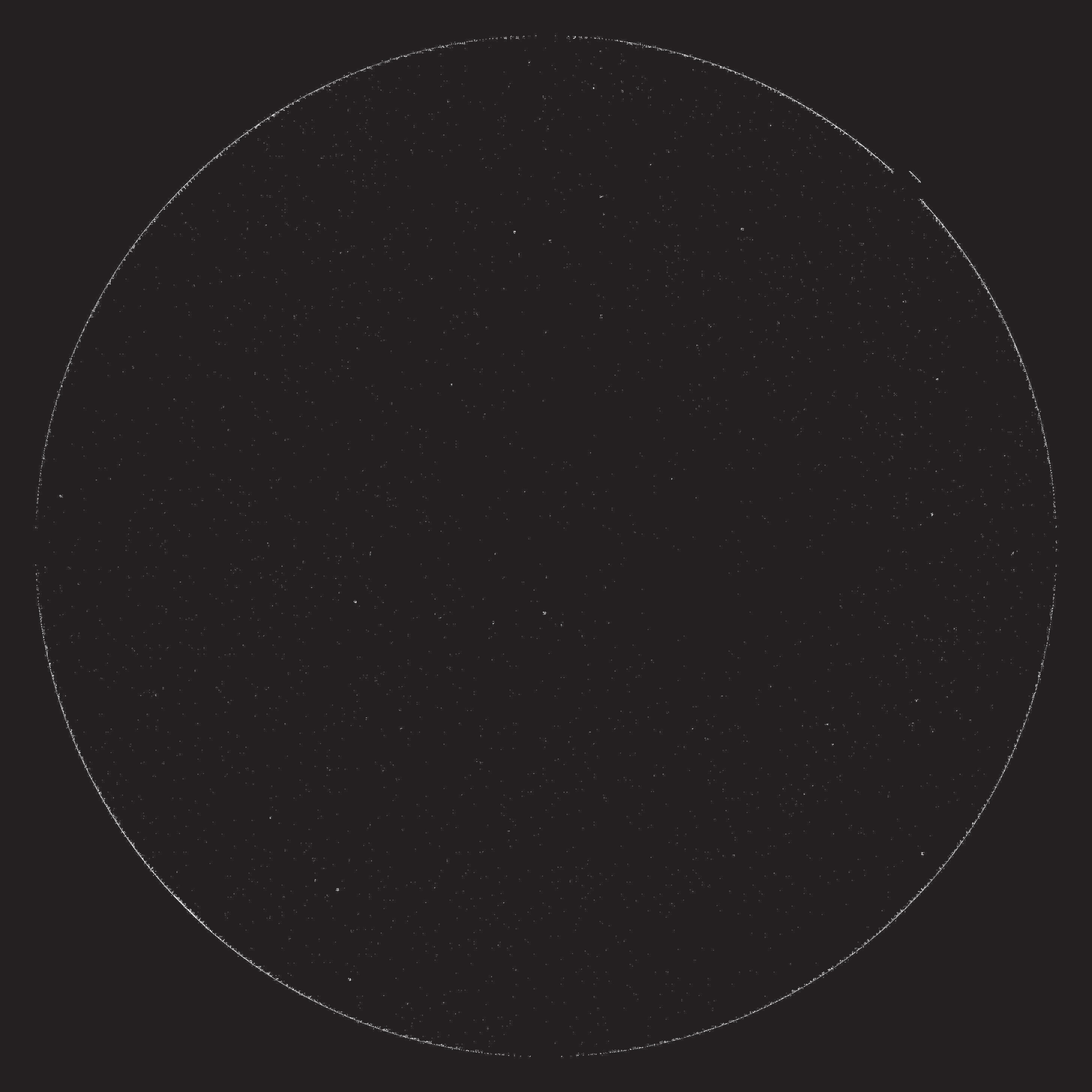}
	}
	\quad
	\subfigure[B-component's detail layer]{
		\includegraphics[width=4cm]{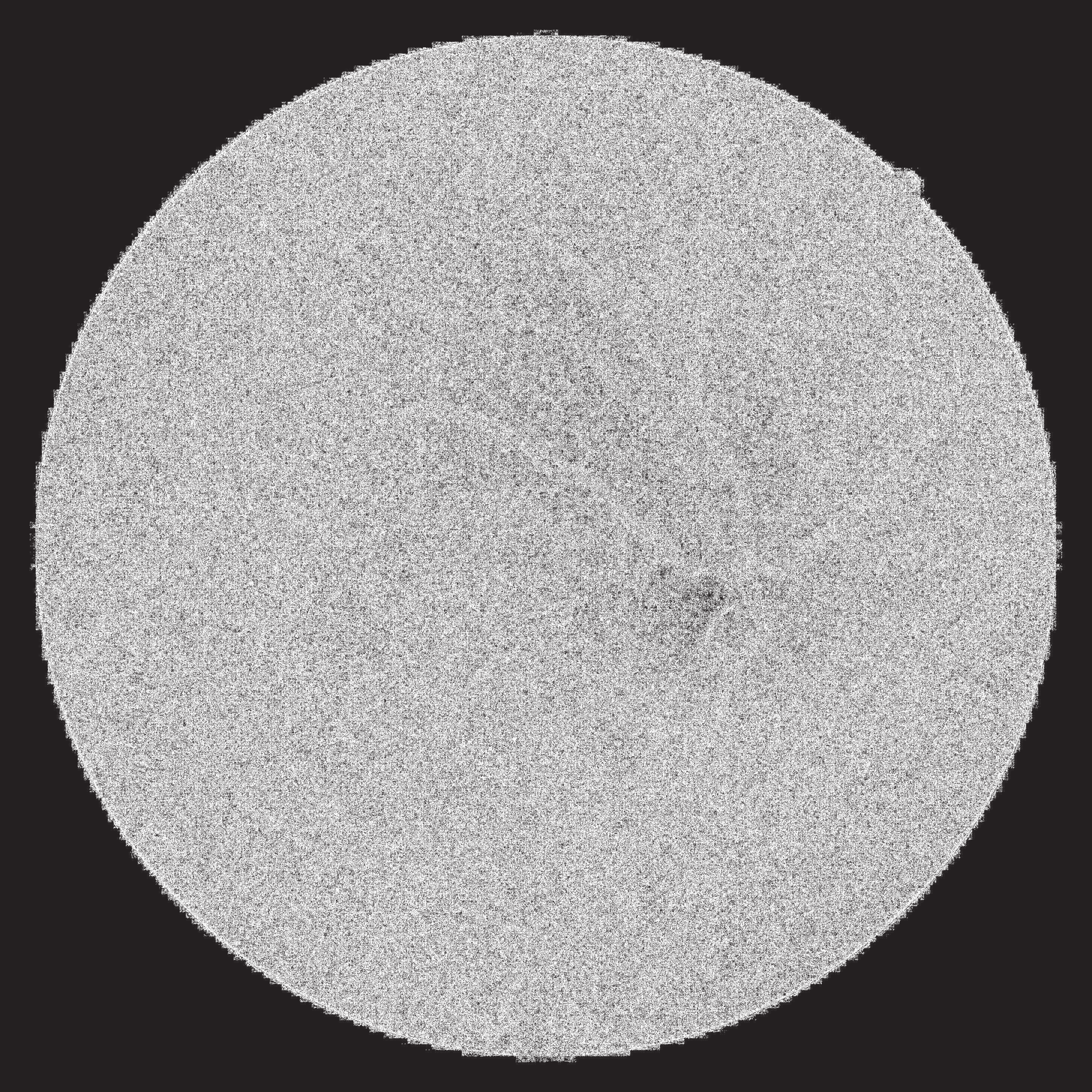}
	}
	\quad
	\subfigure[G-component's shape layer]{
		\includegraphics[width=4cm]{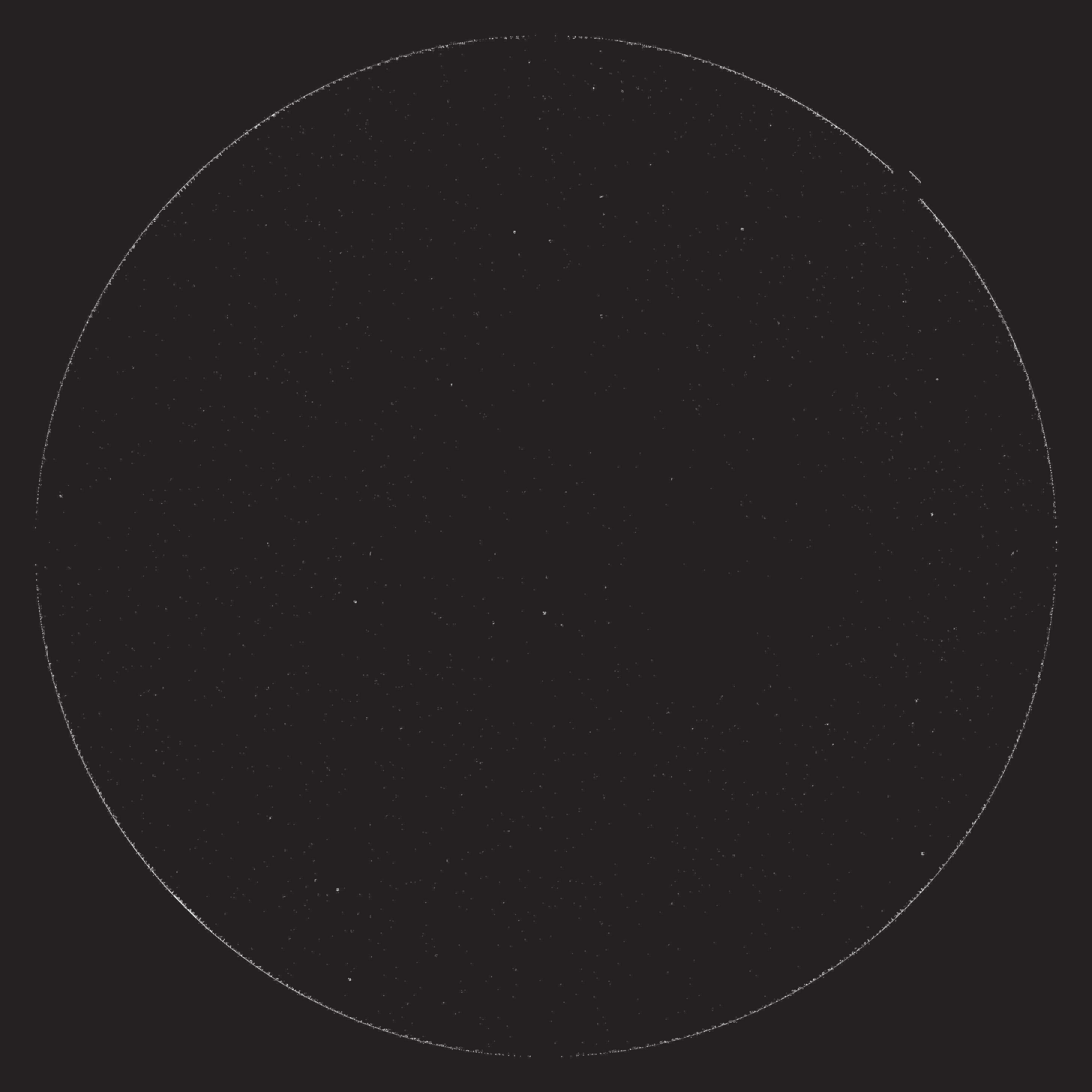}
	}
	\quad
	\subfigure[G-component's detail layer]{
		\includegraphics[width=4cm]{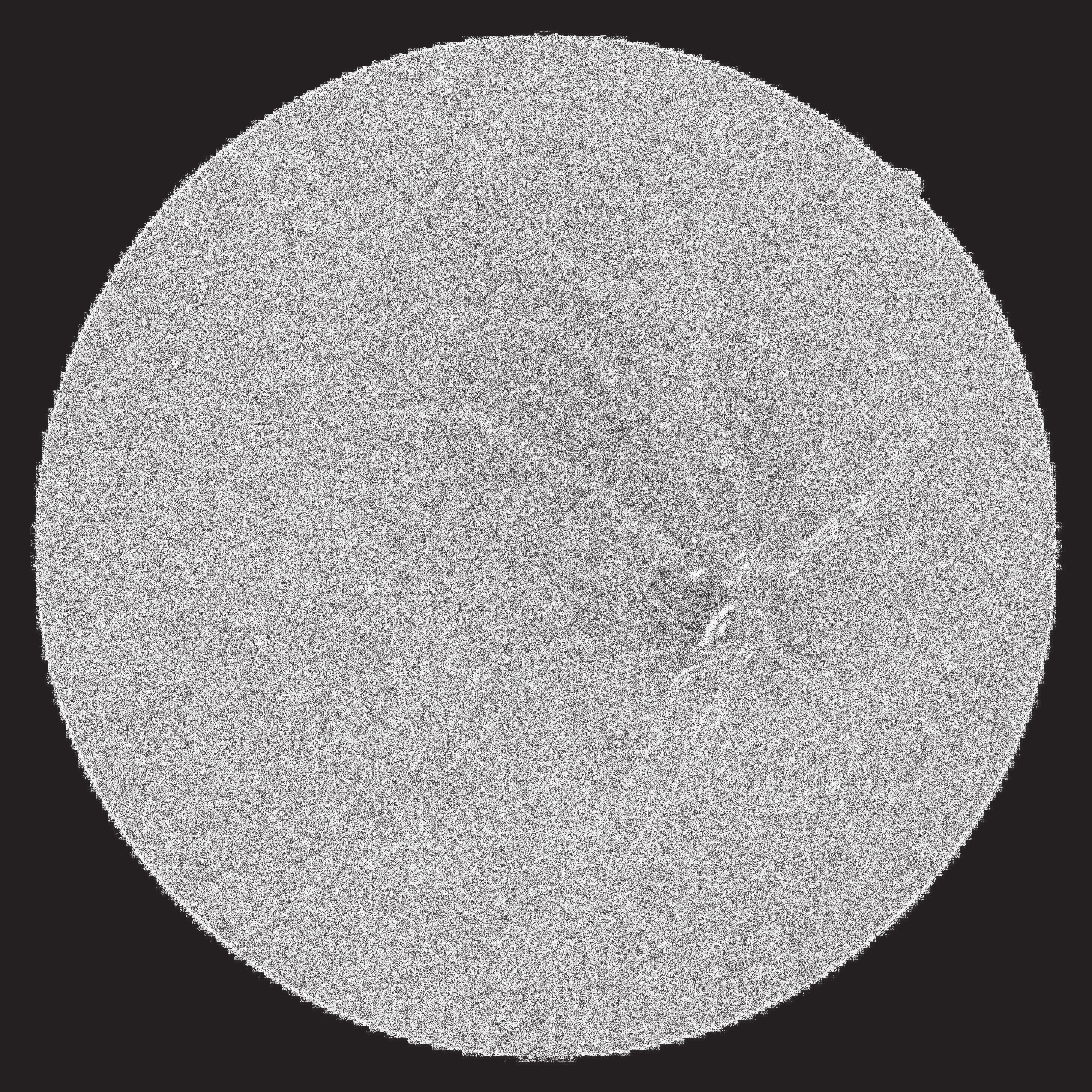}
	}
	\quad
	\subfigure[R-component's shape layer]{
		\includegraphics[width=4cm]{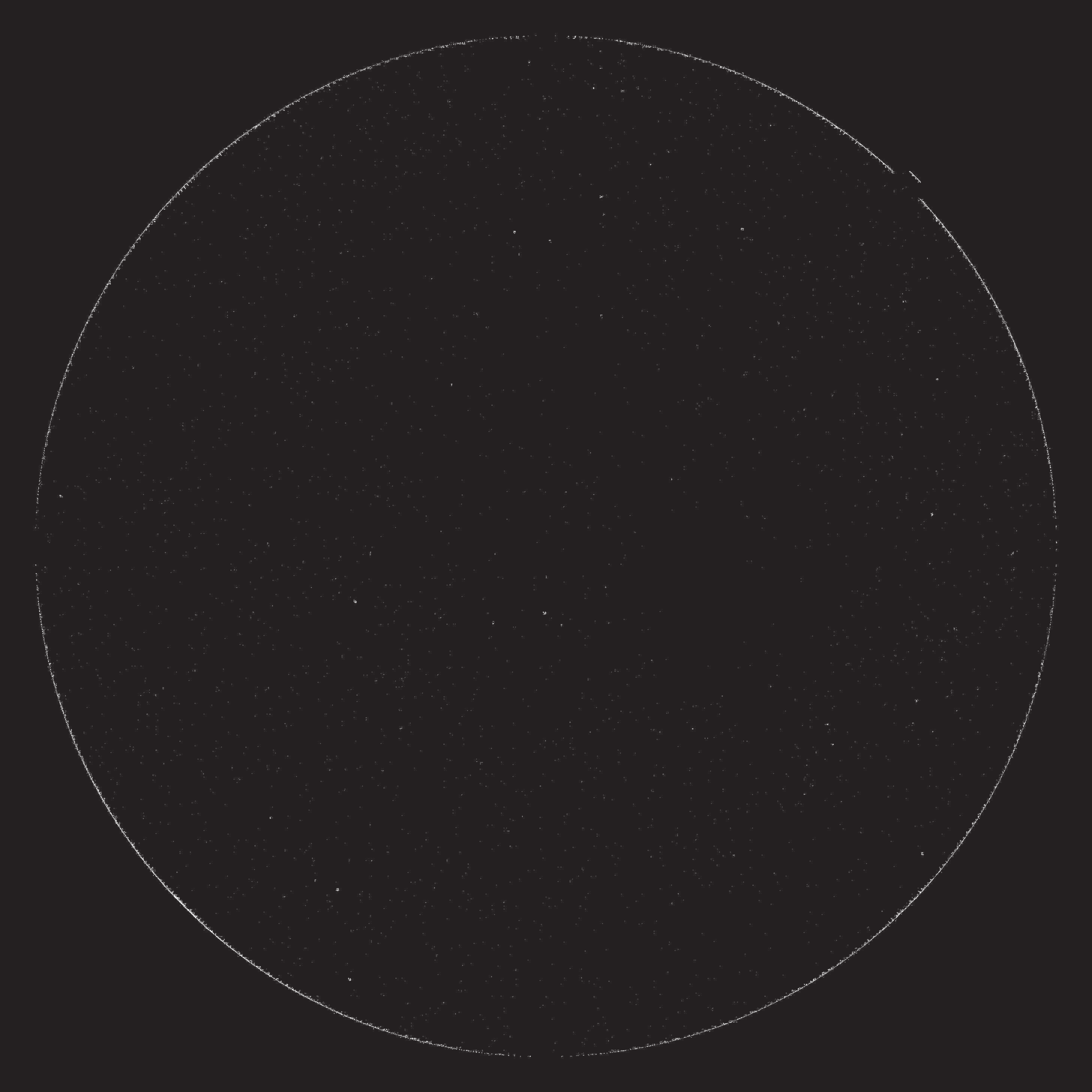}
	}
	\quad
	\subfigure[R-component's detail layer]{
		\includegraphics[width=4cm]{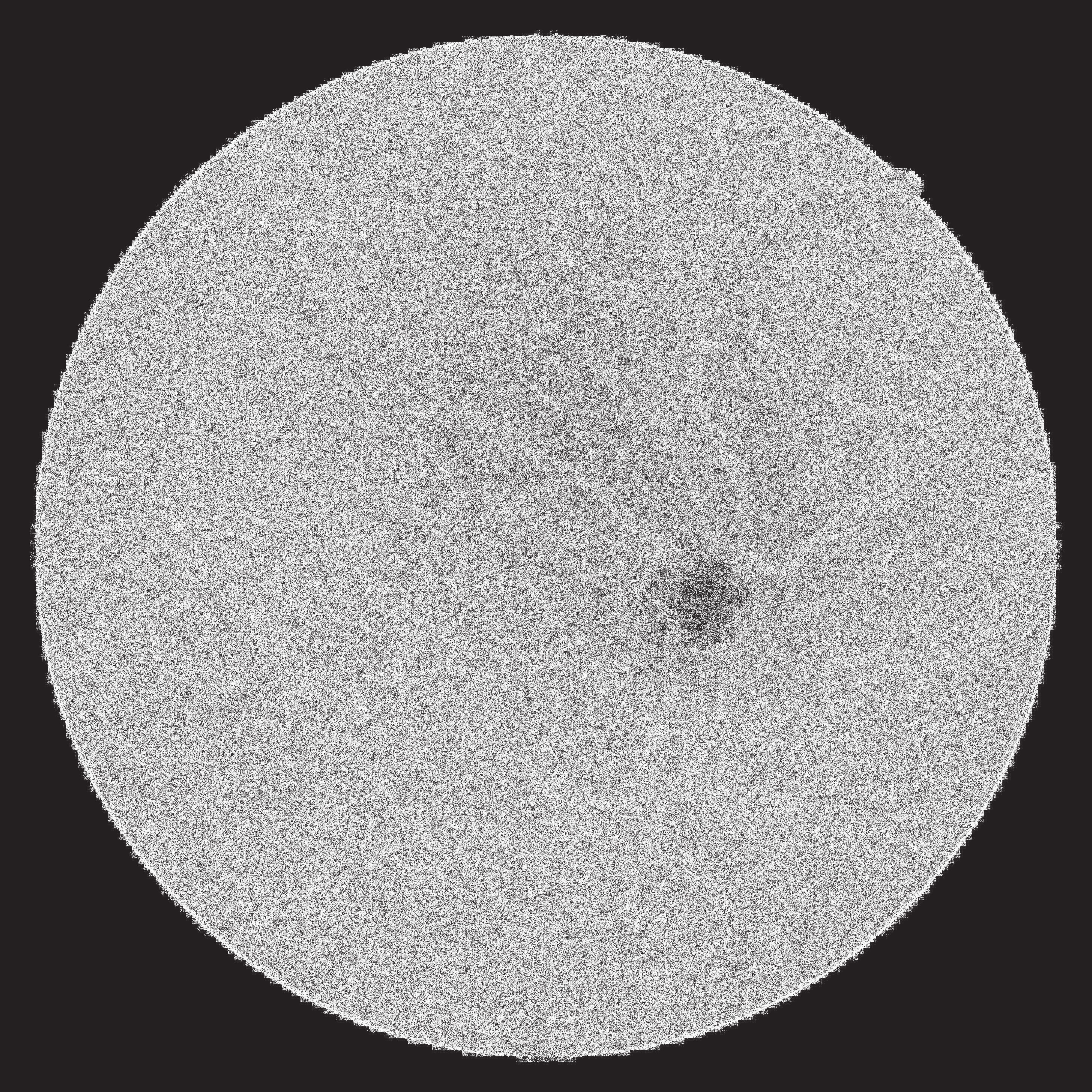}
	}
	\quad
	\subfigure[Reconstructed image]{
		\includegraphics[width=4cm]{1.pdf}
	}
	\caption{An instance of Fundus image registration dataset by using soft compression algorithm for multi-component image}
	\label{Fig:Multi-component}
\end{figure*}

We use ten images in the dataset to train the codebook, and test the compression effect of another image. Fig. \ref{Fig:Multi-component} illustrates an example, whose compression ratio is 3.40. Subfigure (a) is a multi-component original image, whose components are B, G and R, respectively, as shown in (b), (c) and (d). Subfigure (e) to (j) show images after dividing these three components into shape layer and detail layer respectively (binarization has been made for clearer appearance). Subfigure (k) can be obtained through decoding, which is the same as (a).

\section{Conclusion}
In this paper, we investigated how to use soft compression to represent an image, which is in lossless mode. It is based on information theory and statistics to eliminate coding redundancy and spatial redundancy at the same time by using locations and shapes of codebook. Due to the adaptability and completeness of codebooks of sot compression, it can always achieve lossless compression effect for any image and any codebook. 

In theory, we also proposed a new concept, compressible indicator function with regard to image and theoretically analyzed the performance of soft compression, which pointed out the suitable scenarios of soft compression. Compressible indicator function also gives a threshold about the average number of bits required to represent a location by using soft compression.

At the same time, we designed soft compression algorithms for binary image, gray image and multi-component image. These algorithms have been tested on the dataset. Experimental results \footnote{The code of soft compression for gray image is available: https://github.com/ten22one/Soft-compression-algorithm-for-gray-image} indicated that soft compression has sound effects on lossless image compression, especially for images which have a large compressible indicator value.

This paper focuses on lossless compression. However, soft compression can also be combined with other transformation methods, such as wavelet transform. Lossy compression can be realized by using soft compression for the coefficients in the transform domain. Soft compression can also be combined with channel coding to enhance the effect of joint source-channel coding. 

It is expected that this work can have excellent applications in situations where errors cannot be tolerated or where there is significant value, such as the CT image processing for diagnosis and treatment of medical image filed, precious cultural relics, deep space exploration, deep sea exploration, digital libraries and so on.

\ifCLASSOPTIONcompsoc
  \section*{Acknowledgments}
\else
  \section*{Acknowledgment}
\fi

This research was supported by Beijing Natural Science Foundation No.4202030.

\ifCLASSOPTIONcaptionsoff
  \newpage
\fi



%
\bibliographystyle{IEEEtran}
\bibliography{IEEEabrv,soft_compression}

%

\begin{IEEEbiography}[{\includegraphics[width=1in,height=1.25in,clip,keepaspectratio]{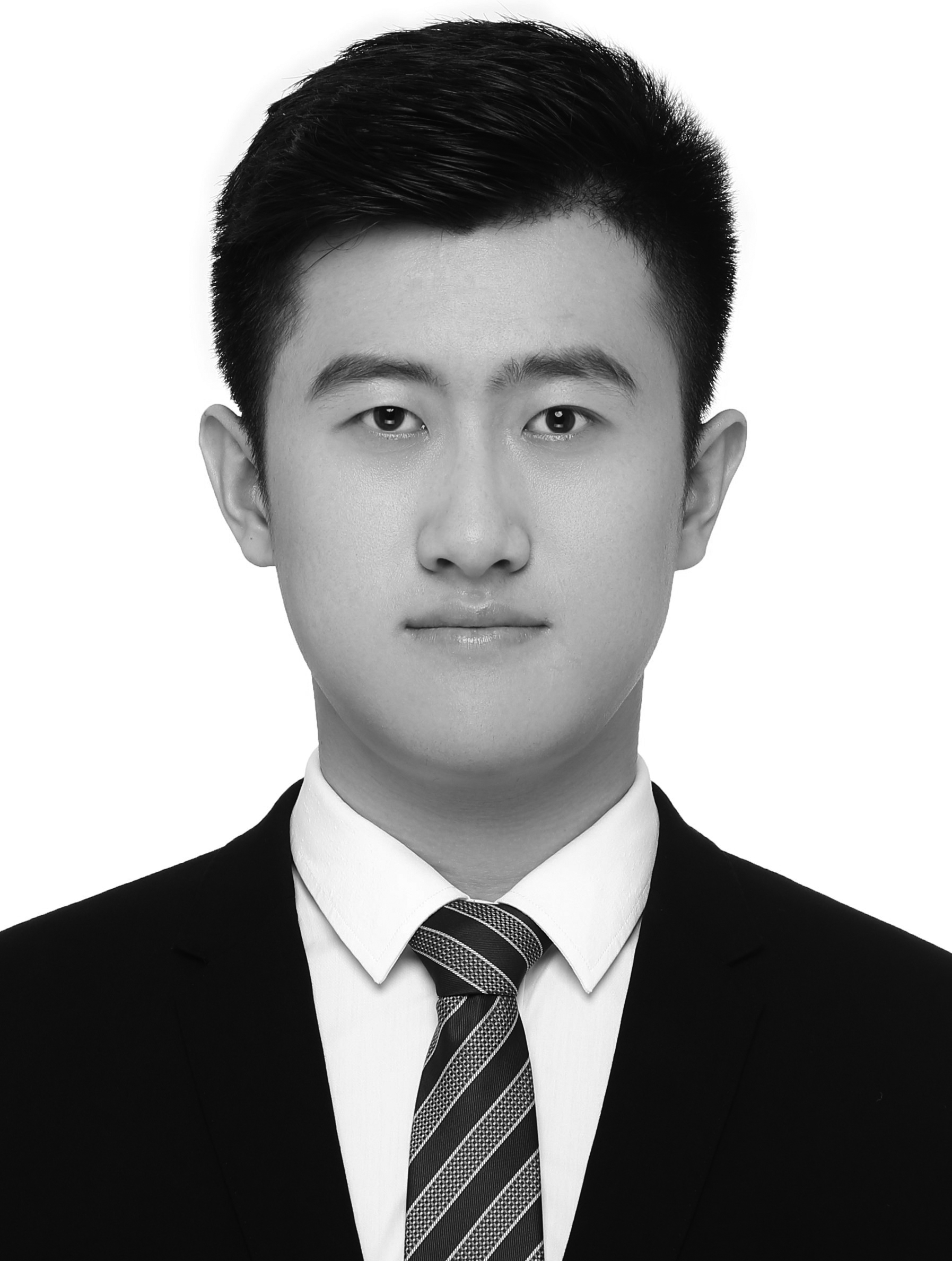}}]{Gangtao Xin}
received the B.S. degree in electronic information science and technology from Nankai University, Tianjin, China, in 2019. He is currently pursuing the Ph.D. degree in electronic engineering with Tsinghua University, Beijing, China. His research interests include image compression and information theory.
\end{IEEEbiography}

\begin{IEEEbiography}[{\includegraphics[width=1in,height=1.25in,clip,keepaspectratio]{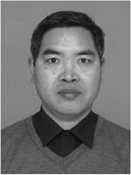}}]{Pingyi Fan}
(M'03--SM'09) received the B.S. degree from the Department of Mathematics, Hebei University, in 1985, the M.S. degree from Nankai University, in 1990, and the Ph.D. degree from the Department of Electronic Engineering, Tsinghua University, Beijing, China, in 1994. From August 1997 to March 1998, he visited the Hong Kong University of Science and Technology as a Research Associate. From May 1998 to October 1999, he visited the University of Delaware, Newark, DE, USA, as a Research Fellow. In March 2005, he visited NICT, Japan, as a Visiting Professor. From June 2005 to May 2017, he visited the Hong Kong University of Science and Technology for many times. From July 2011 to September 2011, he was a Visiting Professor of the Institute of Network Coding, Chinese University of Hong Kong. He is currently a Professor with the Department of EE, Tsinghua University. 

His main research interests include big data analytics, machine learning, 5G technology in wireless communications, such as massive MIMO, OFDMA, network coding, and network information theory. He is an Oversea Member of IEICE. He has attended to organize many international conferences, including as the General Co-Chair of the IEEE VTS HMWC2014, the TPC Co-Chair of the IEEE International Conference on Wireless Communications, Networking and Information Security (WCNIS 2010), and a TPC Member of the IEEE ICC, Globecom, WCNC, VTC, and Infocom. He has received some academic awards, including the IEEE Globecom 2014 Best Paper Award, the IEEE WCNC'08 Best Paper Award, the ACM IWCMC'10 Best Paper Award, the IEEE WCNC'08 Best Paper Award, the ACM IWCMC'10 Best Paper Award, the IEEE ComSoc Excellent Editor Award for the IEEE \textsc{Transactions on Wireless Communications}, in 2009. He has served as an Editor of the IEEE \textsc{Transactions on Wireless Communications}, the \emph{Inderscience International Journal of Ad Hoc and Ubiquitous Computing}, and the \emph{Wiley Journal of Wireless Communications and Mobile Computing}. He is also a Reviewer of more than 32 international journals, including 20 IEEE journals.
\end{IEEEbiography}






\end{document}